\documentclass[epj,final]{svjour}
\usepackage{graphics}
\begin{document}
\title{X--Ray Transitions from Antiprotonic Noble Gases}
\author{D. Gotta\inst{1}\thanks{\emph{corresponding author:} d.gotta@fz-juelich.de}, 
        K. Rashid\inst{2}\thanks{\emph{present address:} 
        National Centre for Physics, QAU campus, Islamabad 44000, Pakistan}, 
        B. Fricke\inst{3}, 
        P. Indelicato\inst{4},
   \and L. M. Simons\inst{5}\thanks{\emph{present address:} 
        Departamento de F{\'i}sica da Universidade de Coimbra, 3004--516 Coimbra, Portugal}
}                     
%
%
\institute{     Institut f\"ur Kernphysik, Forschungszentrum J\"ulich, D--52425 J\"ulich, Germany
           \and Department of Computer Science, Bahria University, Sector E--8, Islamabad 44000, Pakistan 
           \and Institut f\"ur Physik, Universit\"at Kassel, D--34132 Kassel, Germany
           \and Laboratoire Kastler Brossel, UPMC--Paris 6 ENS CNRS; Case 74, 4 place Jussieu, F--75005 Paris, France 
           \and Laboratory for Particle Physics, Paul Scherrer Institut, CH--5232 Villigen, Switzerland}
\date{Received: date / Revised version: date}
%
\abstract{
The onset of antiprotonic X--ray transitions at high principal quantum numbers and the occurence of 
electronic X--rays in antiprotonic argon, krypton, and xenon has been analyzed with the help of 
Multiconfiguration Dirac--Fock calculations. The shell--by--shell ionisation by Auger electron emission, 
characterised by appearance and disappearance of X--ray lines, is followed through the antiprotonic cascade 
by considering transition and binding energies of both the antiproton and the remaining electrons. 
Electronic lines could be attributed partly to specific states of the antiprotonic atom de--excitation.
} 
\authorrunning{D.\,Gotta et al.}
\maketitle
\section{Introduction}
\label{sec:intro}
The capture of an antiproton by atoms into a Coulombic bound orbit and the first 
stages of its subsequent de--excitation in the presence of the remaining electron 
cloud is a highly complicated many--body process. Experimental investigations of the 
upper cascade are difficult because of the small energy gain in the initial steps 
as is the theoretical description because of the interplay of competing processes.

\begin{figure}[b]
\resizebox{0.48\textwidth}{!}{\includegraphics{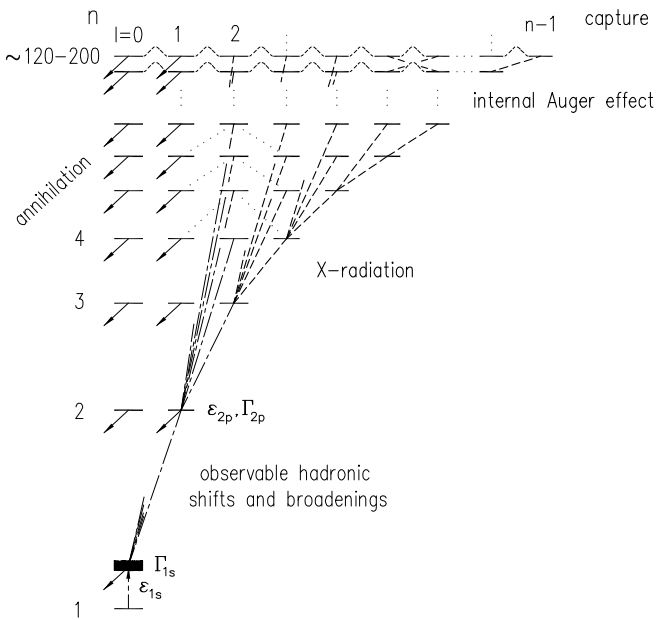}}
\caption{De--excitation cascade in a light antiprotonic atom.}
\label{fig:pbar_fig01_cascade}
\end{figure}

Capture occurs when the antiproton is slowed down to a few tens eV of kinetic energy, 
where collisions, electron excitation and ejection open channels for an antiprotonic 
transition from the continuum into a highly excited quantum state of the Coulomb potential 
of the nucleus. At the beginning of de--excitation the antiproton experiences a 
significantly screened Coulomb force due to the many remaining electrons. Cascading 
through the electron cloud the antiproton knocks out more electrons and rapidly approaches 
the nucleus. After reaching lower--lying levels de--excitation takes place predominantly 
by X--ray emission. Finally the overlap of the orbitals with the nucleus leads to annihilation 
(Fig.\,\ref{fig:pbar_fig01_cascade}). Details on capture and atomic cascade may be found 
in\,\cite{haf74,vog75,vog77,sch78,dan79,egi84,har90,coh04}.

As a rule of thumb the quantum cascade of the captured antiproton starts at large principal 
quantum numbers at about distances of the outermost electron shell. This corresponds to  
$n^{initial}_{\rm{\bar{p}}}=n_{\rm{e}}\sqrt{m_{\rm{\bar{p}}}/m_{\rm{e}}}$, where $n_{\rm{e}}$ is 
the principal quantum number of this electron shell. A first Auger emission at capture may be followed 
by additional electron removal due to shake--off being nearly equally important for all nuclear 
charges\,\cite{zor86}. Calculations suggest that radiative capture is negligible\,\cite{haf74}, 
which has been confirmed experimentally in the case of muonic atoms\,\cite{kae79,har82}. Hence, 
the measurement of X--rays originating from early steps of the electromagnetic cascade provides a way 
to obtain detailed information on the dynamics of the depletion of the electron shells. For antiprotonic 
argon, krypton, and xenon with $n_{\rm{e}}=3$, 4, and 5 for the outermost electron shells,  
$n_{\rm{\bar{p}}}=128$, 170, and 215 allowing a much larger number of de--excitation steps than in 
muonic or pionic atoms. 

Cascade models assume at capture a modified statistical distribution of the angular momentum 
states $\ell $ according to 
$P(n,\ell )\propto (2\ell +1)e^{\alpha \ell }\label{eq:pop},$
where $\alpha$ is a phenomenological parameter obtained from adjusting 
X--ray intensities to the measured line yields~\cite{aky78}. The value of $\alpha = 0$ 
corresponds to a purely statistical distribution. Usually muonic and pionic data are described better 
by modified statistical distributions typically with the modulus of $\alpha$ being less than 0.2, 
but varying from element to element and even with the target chemistry~\cite{har90,har82,fer47,kir99,hor84,shi96}. 
However, it must be emphasized that the population at capture is derived from X--ray transitions 
between low--lying states, i.\,e., it is more or less a reasonable starting point for 
cascade calculations but -- in particular in the case of antiprotons -- far below the initial 
capture states. Theoretical studies suggest that the initial quantum number $n_{\rm{\bar{p}}}$ 
is expected to show a broad distribution but peaked at about $n^{initial}_{\rm{\bar{p}}}$ and 
with different $\ell$ distributions for each $n_{\rm{\bar{p}}}$\,\cite{coh00}. 

For medium and high $Z$ atoms the cascade time from formation to annihilation is of the order of 
picoseconds or less, especially because of the very fast Auger process\,\cite{bur53,deBor54}. 
Therefore, during the antiproton's descent, electrons are continuously removed leading to a depletion 
of electron shells. Significant Auger electron emission has been found in kaonic and muonic atom 
experiments\,\cite{con64a,cal82}.

\begin{figure}[t]
\resizebox{0.45\textwidth}{!}{\includegraphics{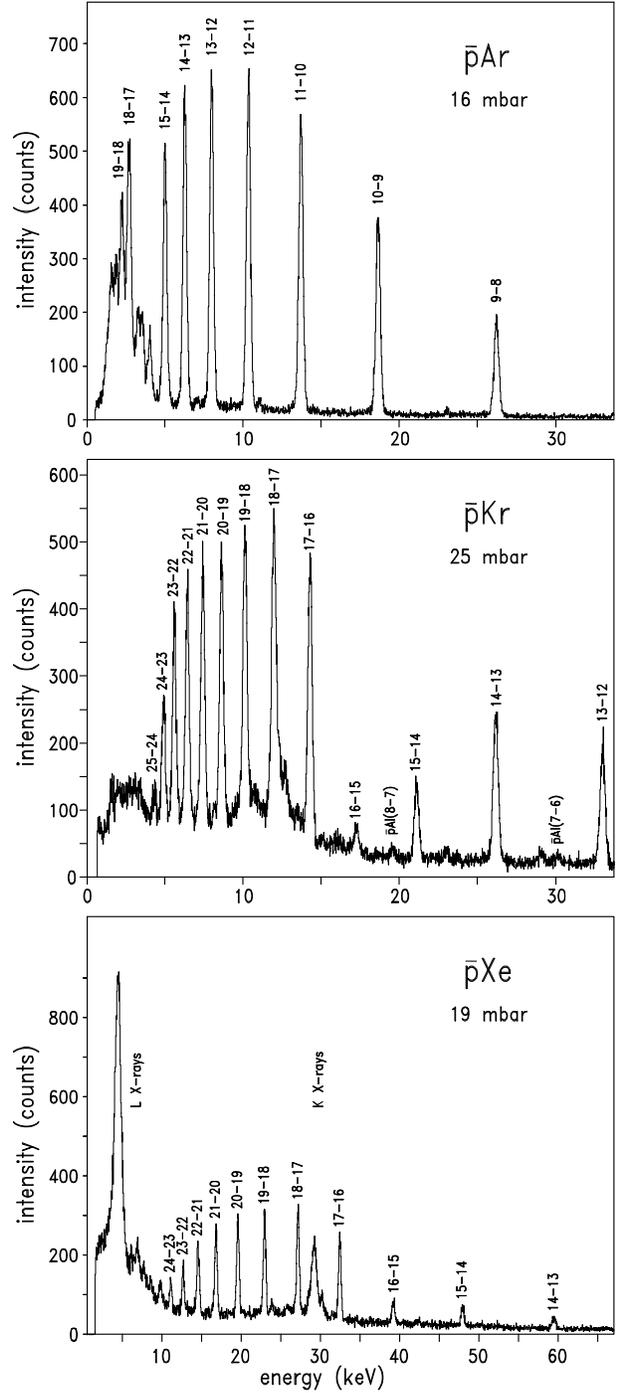}}
\caption{X--ray spectra of antiprotonic argon, krypton, and xenon.}  
\label{fig:Ar_Kr_Xe_spec_full}
\end{figure}

In dilute gases, electron refilling from neighbouring atoms or molecules 
is strongly suppressed during the short cascade time. Consequently, for light and medium--$Z$ atoms 
the number of Auger transitions is sufficient for complete ionisation of the L and higher shells. 
This was established by the observation of X--ray transitions saturated in intensity for elements 
up to krypton along with a reduction of the X--ray yield when the K--electron emission 
threshold is reached at $n_{\rm{\bar{p}}}\approx 16$ (Fig.\,\ref{fig:Ar_Kr_Xe_spec_full}).  
Absolute line yields of the order of 90\% are observed~\cite{bac88,sim94} 
and a distinction of the transition energy for antiprotonic atoms with and without K--shell electrons 
becomes necessary (Table\,\ref{tab:pbar_en_dn1}). 

The frequent Auger emission of electrons during the antiproton's descent causes a continuous and 
successive rearrangement of the electron configurations. Because electrons are much less massive 
than the antiproton their de--excitation occurs immediately whereas atomic antiprotonic states are 
comparatively long--living. While radiative vacancy de--excitation is significant for K holes for krypton 
and higher $Z$, non--radiative de--excitation dominates for L shells up to $Z\approx 90$\,\cite{kes74},  
Therefore, the appearance of electronic L--X--rays yields information on the ionisation state. 

The de--excitation of an exotic--atom cascade does not lead to an excessive vacancy cascade in contrast to 
normal atoms\,\cite{kra67}. It occurs stepwise, because K-- or L--shell Auger emission can take place only 
after the antiproton reaches a sufficiently deep bound state. By then, the outer shells are already 
strongly depleted (see Sec.\,\ref{sec:cascade}).

In the spectra of antiprotonic argon, krypton, and xenon, additional lines appear having 
about the same energies as electronic K and L lines and are in the range of
antiprotonic transitions with $n_{\rm{\bar{p}}}=18 \to 17$ and $n_{\rm{\bar{p}}}=17 \to 16$ 
and around $n_{\rm{\bar{p}}}=30$, respectively (Fig.~\ref{fig:Ar_Kr_Xe_spec_full}). The intensity of these 
lines increases with nuclear charge, i.\,e., with the number of electrons. In the lightest elements 
like neon, such electronic radiation falls below the detection threshold as is the case for the 
L series in argon.

X---ray emission in general is expected to increase \linebreak 
strongly with nuclear charge $Z$ because of the increasing fluorescence. Therefore, electronic X--rays 
have been observed mainly in heavy muonic and pionic atoms and in some cases from their energy shifts 
information on the ionisation state has been deduced\,\cite{har82,arl74,fro75,sch80,nin05}. Significant 
or even complete depletion of the electron shells, however, is only possible for light nuclei together 
with dilute targets\,\cite{kir99,bac85,hau98}. In solid targets, in particular in the case of metals, 
due to instant refilling the exotic--atom cascade proceeds more or less in the presence of the electron 
shells as seen from a K$\alpha$--to--K$\beta$ ratio as expected from an electronic $Z-1$ atom\,\cite{har82}.

\begin{table}[t]
\caption{Transition energies between circular states in antiprotonic atoms argon, krypton, and xenon atoms, which 
 are electron--free and with 2 remaining K electrons.
 The first two antiprotonic $\Delta n_{\rm{\bar{p}}}=1$ transitions, which are able to ionise the
 K--shell electrons are printed bold face.}
\label{tab:pbar_en_dn1}
\setlength{\tabcolsep}{1.9mm}
\begin{tabular}{ccccccc}
\hline\noalign{\smallskip}
$Z$                                    &  18    &  18  &   36       &  36  &   54      & 54\\
$n_{\bar{p}}$                          & $\rm{\bar{p}}$Ar& $\rm{\bar{p}}$Ar$2e^{-}$&$\rm{\bar{p}}$Kr&$\rm{\bar{p}}$Kr$2e^{-}$&$\rm{\bar{p}}$Xe&$\rm{\bar{p}}$Xe$2e^{-}$\\
$\rightarrow$                          & \multicolumn{6}{c}{} \\
$n_{\bar{p}}-1$                        & \multicolumn{6}{c}{energy (eV)} \\
\hline
\end{tabular}
\setlength{\tabcolsep}{1.9mm}
\begin{tabular}{crrrrrr}
8--7                                   & 37193       &      &            &      &           &\\
9--8                                   & 25972       &      &            &      &             &\\
10--9                                  & 18566       &      &            &      &             &\\
11--10                                 & 13729       &      &            &      &             &\\
12--11                                 & 10438       &      &            &      &             &\\
13--12                                 & 8120        &      &32965       &      &             &\\
14--13                                 & 6441        &      &26145       &      & 59167       &\\
15--14                                 & 5196        &      &{\bf 21084} &      & {\bf 47708} &\\
16--15                                 & {\bf 4251}  & 4249 &{\bf 17250} &      & {\bf 39028} & 39018\\
17--16                                 & {\bf 3523}  & 3520 &14292       &14286 & 32333       & 32321\\
18--17                                 & 2952        & 2949 &11974       &11967 & 27086       & 27073\\
19--18                                 & 2498        & 2495 &10130       &10123 & 22916       & 22901\\
20--19                                 & 2133        & 2129 &8649        &8639  & 19560       & 19543\\
21--20                                 & 1835        & 1831 &7442        &7431  & 16829       & 16810\\
22--21                                 & 1591        & 1586 &6449        &6438  & 14584       & 14563\\
23--22                                 &             &      &5626        &5613  & 12721       & 12698\\
24--23                                 &             &      &4937        &4923  & 11163       & 11138\\
25--24                                 &             &      &4356        &4341  & 9849        & 9822\\
26--25                                 &             &      &3863        &3847  & 8733        & 8704\\
27--26                                 &             &      &3442        &3424  & 7780        & 7750\\
28--27                                 &             &      &3079        &3061  & 6961        & 6928\\
29--28                                 &             &      &2766        &2746  & 6253        & 6219\\
30--29                                 &             &      &2494        &2473  & 5638        & 5602\\
31--30                                 &             &      &2257        &2235  & 5101        & 5063\\
32--31                                 &             &      &2048        &2025  & 4630        & 4591\\
33--32                                 &             &      &1865        &1841  & 4215        & 4174\\
34--33                                 &             &      &1703        &1678  & 3849        & 3806\\
35--34                                 &             &      &1558        &1532  & 3523        & 3478\\
36--35                                 &             &      &            &      & 3233        & 3186\\
37--36                                 &             &      &            &      & 2974        & 2925\\
38--37                                 &             &      &            &      & 2743        & 2693\\
\hline\noalign{\smallskip}
\end{tabular}
\end{table}

In this paper we discuss the case of the noble gases argon, krypton, and xenon, i.\,e., for 
electron shells originally occupied with 18, 36, and 54 electrons, the possible origin of the additional 
transitions in the energy range of fluorescence X--rays and their relation to the antiprotonic state   
during the atomic cascade. At pressures around 20~mbar the collision rate for, e.\,g., thermalised 
$\rm{\bar{p}}$Ar is estimated to about $3\cdot 10^6$/s, whereas cascade times are $10^{-9}$\,s or less. 
So electron refilling is strongly suppressed. 

A short discussion of electronic X--rays emitted during 
the antiprotonic cascade was given by Bacher\,\cite{bac88,bac87} and Simons\,\cite{sim94}. For the present 
analysis higher statistics measurements have been used and the electronic and antiprotonic 
binding energies have been calculated in Multiconfiguration Dirac-Fock (MCDF) approximation including
radiative and nuclear finite size corrections by using the code of Indelicato and Desclaux\,\cite{MCDF}.

For the low--lying atomic levels strong interaction leads to a state dependent energy shift
and level broadening. The highest circular states ($n_{\rm{\bar{p}}},\ell_{\rm{\bar{p}}} = n_{\rm{\bar{p}}}-1$) 
affected by annihilation of antiprotons for atoms with $Z=10,18,36$, and 54 are 
$(n_{\rm{\bar{p}}},\ell_{\rm{\bar{p}}})=(4,3),(5,4),(7,6)$, and(8,7), respectively\,\cite{pot79}. 
Transitions affected by strong interaction are not considered here.

\section{Experiment}
\label{sec:exp}

The X--ray spectra were measured at an extracted beam of the Low--Energy Antiproton Ring
(LEAR) at CERN by using the set--up of LEAR experiment PS175, the main goal of which 
was the determination of strong--interaction effects in antiprotonic hydrogen and 
helium isotopes formed at low densities\,\cite{PS175,bac89a,sch91,hei92}.
The set--up, which included the cyclotron trap, allowed to stop up to 90\% of the 105\,MeV/c
antiproton beam in a gaseous target of a few tens of mbar pressure in
a volume of about 20\,cm$^{3}$ only. The X-rays were detected with a Si(Li) detector
of 30\,mm$^{2}$ inner area surrounded by a 200\,mm$^{2}$ outer area, which was
used in anticoincidence for background reduction (guard--ring configuration). The detector
was located 16\,cm away from the stop volume. In--beam energy resolution of the detector 
was measured to be 280\,eV at 6.4\,keV. Details about the experimental set--up may be found 
in\,\cite{bac89a,sch91,hei92}.

\begin{figure}[h]
\resizebox{0.45\textwidth}{!}{\includegraphics{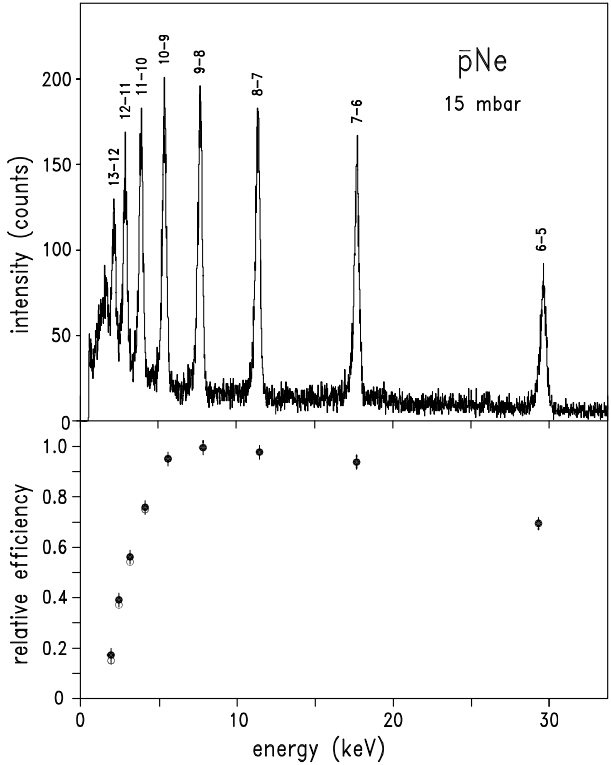}}
\caption{Top -- Antiprotonic neon as measured with a 30/200\,mm$^{2}$ Si(Li) guard--ring detector.
           The relative detector efficiency is given by the line intensities because of the 
           saturated line yields. Bottom -- Relative detection efficiency as derived from the saturated 
           lines of the $\rm{\bar{p}}$Ne spectrum. The upper points (full symbols) are 
           due to the correction of self absorption of X--rays in neon.}
\label{fig:pbar_Ne_rel_eff}
\end{figure}

In--beam detection efficiency and resolution have been determined from the saturated circular 
transitions of fully stripped low--$Z$ antiprotonic systems like $\rm{\bar{p}}$Ne\,\cite{bac88,sim94}
up to an energy of 30\,keV (Fig.\,\ref{fig:pbar_Ne_rel_eff}). For higher energies the efficieny is 
calculated from the conversion probability in 3.5\,mm silicon given by the photo cross section\,\cite{vei73}.  
A general feature of all X--ray spectra is an increasing background towards lower energies 
caused by the electromagnetic showers induced by the annihilation products in the surrounding materials.

\section{Antiprotonic cascade in the presence of electrons}
\label{sec:cascade}

The initial de--excitation of the captured antiproton is dominated by
$\Delta n_{\rm{\bar{p}}} = 1$ E1 Auger emission of s electrons whenever energetically 
possible\,\cite{bur53,deBor54}. The general behaviour is easily deduced from an inspection 
of Ferell's formula\,\cite{fer60}
\begin{eqnarray}
\it\Gamma_{A}/\it\Gamma_{X}=\sigma_{\gamma e}^{(Z^{e}_{eff}-1)}(E)/[(Z^{e}_{eff}-1)^{2}\sigma_{T}],
\end{eqnarray}
which relates the ratio of Auger emission and X--ray transition rate to the energy dependence 
of the photoelectric cross section (Fig.\,\ref{fig:pbar_fig04_GA_GX_Gtot}). Here, 
{$\sigma_{\gamma e}(E)$ and $\sigma_{T}$ denote the photoelectric and the Thomson cross sections. 
$Z^{e}_{eff}$ is the effective charge seen by the electron. Reducing $Z^{e}_{eff}$ by one unit 
takes into account the screening caused by the antiproton's charge. 

\begin{figure}[b]
\resizebox{0.45\textwidth}{!}{\includegraphics{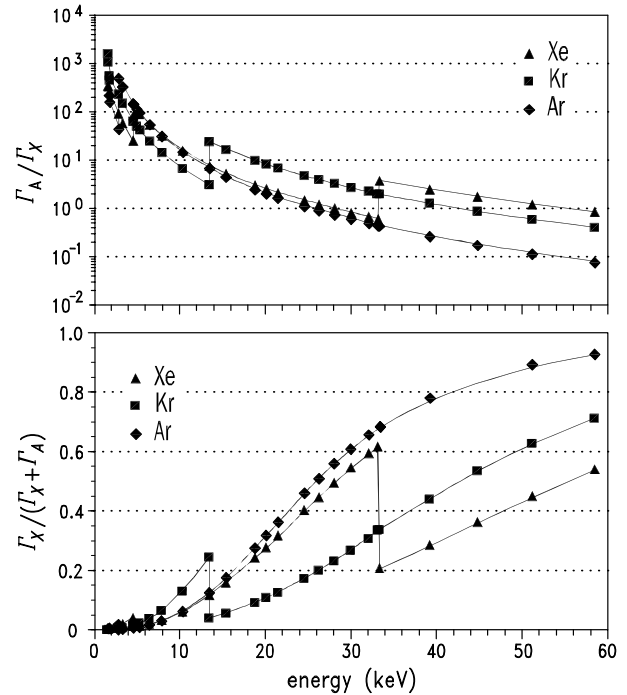}}
\caption{Top -- relative strength of Auger to radiative de--excitation as calculated from
         Ferell's formula for fully occupied electron shells\,\protect\cite{bac88}.
         Here $Z^{e}_{eff}=Z-1$ is taken, which corresponds to an antiproton already 
         inside the electronic K shell. 
         Bottom -- relative strength of X--ray to total de--excitation. In the case the 
         electron's binding energy is larger than the energy gain of the antiprotonic 
         transition, Auger emission is suppressed.}
\label{fig:pbar_fig04_GA_GX_Gtot}
\end{figure}

Because the radiative decay width $\Gamma_{X}\propto\Delta E^{3}$ has a strong
energy dependence, $\Delta n_{\rm{\bar{p}}}>1$ radiative transitions from low  angular
momentum states $\ell$ compete with Auger de--excitation. Hence, radiative transitions tend to 
populate circular states ($n_{\rm{\bar{p}}},\ell = n_{\rm{\bar{p}}}-1$), whereas 
$\Delta n_{\rm{\bar{p}}}=1$ Auger transitions leave the distribution of angular momentum states
essentially unchanged. For large initial values $n_{\rm{\bar{p}}}$ the energy gain for the 
$\Delta n_{\rm{\bar{p}}} = 1$ transition favoured by the Auger process ($\Gamma_{A}\propto 1/\surd{E}$) 
is not sufficient for ionisation. Hence, it is expected that $\Delta n_{\rm{\bar{p}}}>1$ Auger 
transitions are also important.

As the Auger process depletes the electronic shells, the reduced screening of the nuclear charge 
increases the binding energy of the remaining electrons. Whenever the binding energy becomes larger 
than the antiprotonic transition energy, Auger de--excitation is interrupted (\textit{effective stoppage of 
energy loss:}\,\cite{fer47}) or even stopped when all atomic electrons are ejected. Hereafter the exotic 
atom looses energy only through $\Delta n_{\rm{\bar{p}}}\geq 1$ radiative transitions, enhancing further 
the population of states with $(n_{\rm{\bar{p}}},n_{\rm{\bar{p}}}-1)$ from where the cascade can continue 
only via circular transitions $(n_{\rm{\bar{p}}},\ell_{\rm{\bar{p}}}=n_{\rm{\bar{p}}}-1)\to
(n_{\rm{\bar{p}}}-1,\ell_{\rm{\bar{p}}}=n_{\rm{\bar{p}}}-2)$. The higher weight of the states with 
large angular momentum $\ell$ in addition enhances the development of a circular cascade. 
For $Z=2$ such a limitation to radiative de--excitation between high--lying circular states leads to 
metastability\,\cite{con64} because of suppression of $\Delta n_{\rm{\bar{p}}},\Delta \ell_{\rm{\bar{p}}}>1$ 
Auger transitions\,\cite{rus70} and is observed by a significantly increased cascade time of a 
fraction of the formed $\rm{\bar{p}}$He atoms\,\cite{yam02}.

The large initial value of $n_{\rm{\bar{p}}}$ allows a large number of de--excitation steps, which leads 
very soon to an almost circular cascade and, up to krypton, to a complete removal of all electrons 
if refilling is avoided\,\cite{bac88}. Hence, in dilute targets low and medium--$Z$ antiprotonic atoms 
exist as true hydrogen--like systems during the intermediate cascade. In the case of xenon, however, 
the non--saturation of the X--ray yields for initial states $n_{\rm{\bar{p}}}\leq 14$ indicates that 
the number of steps is not sufficient for complete removal of 54 electrons. Details are discussed in 
Sec.\,\ref{sec:cascade}.
 
We characterize the complex interplay between the antiproton and the electron shells by the antiprotonic 
quantum number $n_{\rm{\bar{p}}}$, the number of electrons present at that stage, and their binding energies. 
When discussing a cascade dominated by circular transitions, i.\,e., $\ell_{\rm{\bar{p}}}=n_{\rm{\bar{p}}}-1$,   
and in addition $\ell_{\rm{\bar{p}}}\gg 1$, the antiproton's probability density is finite only around a 
rather well defined distance $R_{\rm{\bar{p}}}$ from the nucleus. Its binding energy is 
determined by $n_{\rm{\bar{p}}}$ and the effective nuclear charge $Z^{\rm{\bar{p}}}_{eff}$, where 
$Z^{\rm{\bar{p}}}_{eff}$ depends on both the distance $R_{\rm{\bar{p}}}$ and on the number of remaining 
electrons, which in turn is related to the level reached by the antiproton. Therefore, the quantities 
$n_{\rm{\bar{p}}}$, $Z^{\rm{\bar{p}}}_{eff}$, $R_{\rm{\bar{p}}}$, and number of electrons are strongly correlated.

To obtain a qualitative description of this correlation, the 
dependence of $Z^{\rm{\bar{p}}}_{eff}$ on the root mean square radius $R_{\rm{\bar{p}}}$ is computed 
for the electronic K, L, M, N, and O shells by using the MCDF code. From these values a continuous function 
$Z^{\rm{\bar{p}}}_{eff}(R_{\rm{\bar{p}}})$ is constructed by interpolation. We obtain for argon, krypton, and xenon
\begin{eqnarray}
Z^{\rm{\bar{p}Ar}}_{eff}(R_{\rm{\bar{p}}}) &=& 18.174 e^{-0.534\cdot R_{\rm{\bar{p}}}}\label{eq:R_pbar_Ar},\\  
Z^{\rm{\bar{p}Kr}}_{eff}(R_{\rm{\bar{p}}}) &=& 36.167 e^{-0.579\cdot R_{\rm{\bar{p}}}}\label{eq:R_pbar_Kr}, and\\  
Z^{\rm{\bar{p}Xe}}_{eff}(R_{\rm{\bar{p}}}) &=& 54.237 e^{-0.8654\cdot R_{\rm{\bar{p}}}+0.002287\cdot R^2_{\rm{\bar{p}}}+0.069956\cdot R^3_{\rm{\bar{p}}}}.
\label{eq:R_pbar_Xe}
\end{eqnarray}
To calculate the (non--relativistic) transition energies \linebreak 
$\Delta E_{n_{\rm{\bar{p}}}\rightarrow n_{\rm{\bar{p}}}-1}$ we start at a value of 
$R_{\rm{\bar{p}}}$ that corresponds to the outermost shell of electrons, where about the 
quantum cascade is expected to start. The corresponding radial quantum number $n_{\rm{\bar{p}}}$ is then given by 
\begin{eqnarray}
n_{\rm{\bar{p}}}&=&\sqrt{R_{\rm{\bar{p}}}\cdot Z^{\rm{\bar{p}}}_{eff}\cdot m_{\bar{p}}/m_{e}}\label{eq:n_pbar}
\end{eqnarray}
and the antiprotonic transition energy for $n_{\rm{\bar{p}}}\rightarrow n_{\rm{\bar{p}}}-1$ by
\begin{eqnarray}
\Delta E_{n_{\rm{\bar{p}}}\rightarrow n_{\rm{\bar{p}}}-1}
&=&E_{Ryd}\cdot \frac{(Z^{\rm{\bar{p}}}_{eff})^2(2n_{\rm{\bar{p}}}-1)}{[n_{\rm{\bar{p}}}(n_{\rm{\bar{p}}}-1)]^2}\label{eq:dE_pbar}
\end{eqnarray}
with the antiprotonic Rydberg energy $E_{Ryd}=m_{\rm{\bar{p}}}\,c^2\alpha ^2/2=24982.2~eV.$
$R_{\rm{\bar{p}}}$ for $n_{\rm{\bar{p}}}-1$ is calculated from equation\,(\ref{eq:n_pbar}), which 
yields the corresponding $Z^{\rm{\bar{p}}}_{eff}$ by using the expressions (\ref{eq:R_pbar_Ar})--(\ref{eq:R_pbar_Xe}).
Inserting the new values for $Z^{\rm{\bar{p}}}_{eff}$ the energy of the next lower transition is obtained from  
(\ref{eq:dE_pbar}). This procedure is repeatedly applied. The approximations made here are regarded to 
be sufficient in view of the complexity of the cascade processes. The results are given in 
Tables\,\ref{tab:pbar_Ar_electrons}\,--\,\ref{tab:pbar_Xe_electrons} along with the electronic 
binding energies for the successive states of ionisation. For each pair of 
$n_{\rm{\bar{p}}}$ and $Z^{\rm{\bar{p}}}_{eff}$ the relativistic transition energy is then calculated 
with the MCDF code. The large mass of the antiproton leads to much smaller radii and, consequently, 
the effect of electron screening is reduced for the observable transitions as compared 
to muonic or pionic atoms\,\cite{vog73,fri75,vog77a}.

In Fig.\,\ref{fig:pbar_fig05_Zeff_R(Zeff)} the results for antiprotonic transition, electronic binding 
energy, and antiproton and electron orbits are combined. One finds that the maximum probability density of 
antiproton and electron do not overlap when ionization occurs by $\Delta n_{\rm{\bar{p}}}=1$ transitions. 
The antiproton has to be always far inside the corresponding electron shell. 

\begin{figure}[h]
\resizebox{0.45\textwidth}{!}{\includegraphics{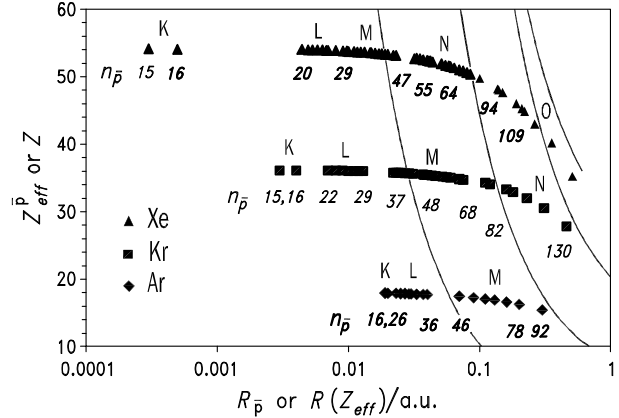}}
\caption{Effective nuclear charge $Z^{\rm{\bar{p}}}_{eff}$ acting on the antiproton as a function 
         of its distance $R_{\rm{\bar{p}}}$. K, L, M, N, and O indicate the regions where these 
         shells can be ionised by antiprotonic $\Delta n_{\rm{\bar{p}}}=1$ transitions. Solid lines 
         show the average radii of the electron shells K, L, M1--M3, an M4--M5 (from left) 
         for the screened nuclear charge $Z$. The different screening for the various subshells is taken into 
         account by the shielding constants $Z_{shell}^{S}$ ($Z_{K}^{S}=0.3$, $Z_{L}^{S}=4.15$, 
         $Z_{M1-M3}^{S}=11.25$, and $Z_{M4-M5}^{S}=21.15$\,\cite{han75}.)}
\label{fig:pbar_fig05_Zeff_R(Zeff)}
\end{figure}

Vice versa, the presence of the antiproton inside the electron cloud reduces the electronic binding 
energies. Because of its large mass the orbit is rather well localised for $\ell \gg 1$ states and, 
therefore, it shields the nuclear charge by almost one unit for all electron shells outside its orbit. 
Binding energy and effective charge $Z^{e}_{eff}$ for varying number of electrons for 
argon, krypton, and xenon are given in Tables\,\ref{tab:Zeff_Ar}\,--\,\ref{tab:Zeff_Xe}. 
With the use of these tables, we shall describe the main features of the antiprotonic cascade in 
the presence of electrons. Although the qualitative picture is similar for these three elements, a
quantitative description shows that differences observed can be attributed to the increasing number 
of electrons. To keep the following discussion transparent, we develop the features for a purely 
circular antiprotonic cascade.

\subsection{Antiprotonic Cascade in Argon (Z=18)}\label{chap:ar_cascade}
\label{subsec:Ar_cascade}

Assuming capture in the spatial region of the 3s and 3p electrons, the effective charge seen by the 
antiproton is about $Z^{\rm{\bar{p}}}_{eff}=8$ and its binding energy is about 16\,eV being equal to the
ionisation potential of the most weakly bound electrons. According to the thumb rule with $n_{e}=3$ the 
radial quantum number for the start of the quantum cascade reads $n_{\rm{\bar{p}}}\approx 43\cdot n_{e}=128$. 
For $n_{\rm{\bar{p}}}\approx 128$ the antiprotonic transition $n_{\rm{\bar{p}}}\rightarrow n_{\rm{\bar{p}}}-1$ 
has an energy of about 5\,eV only, whereas the binding energies of the 3s, 3p$_{1/2}$, and 3p$_{3/2}$ 
electrons are 35, 16.2 and 16.0\,eV, respectively (Table\,\ref{tab:Zeff_Ar}). Consequently the antiproton 
must descend the cascade ladder by radiative transitions until it gains enough energy to remove the 
M electrons or the cascade has to start at somewhat lower--lying states. Figure~\ref{fig:argon_en_pbar_Zeff} 
shows the range for the quantum number $n_{\rm{\bar{p}}}$, where Auger emission is allowed in 
$\Delta n_{\rm{\bar{p}}}=1 $ transitions for the various electron shells.

\begin{figure} [t]
\resizebox{0.48\textwidth}{!}{\includegraphics{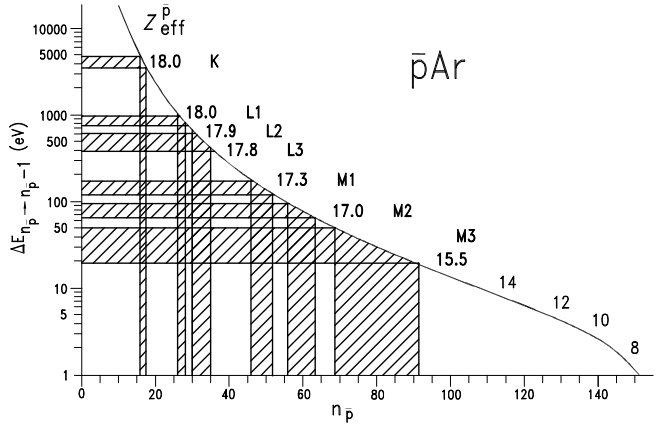}}
\caption[Energies of $\Delta n_{\rm{\bar{p}}}=1$ transitions in antiprotonic argon] 
 {Energies of circular $\Delta n_{\rm{\bar{p}}}=1$ transitions in antiprotonic argon.
  The effective charge $Z_{\rm{eff}}$ acting on the antiproton determines
  when depletion of the various electron shells starts. In the beginning, $Z_{\rm{eff}}$
  varies strongly with the number of remaining electrons being about 15 when 
  M--shell emission starts. For L--shell emission, the antiproton
  is already closer to the nucleus than any electron. Hence, $Z_{\rm{eff}}$
  approaches the value of the unscreened nuclear charge of 18.}
\label{fig:argon_en_pbar_Zeff}
\end{figure}

\begin{table*}[t]
\caption
 {State and cascade dependent ionisation energies in antiprotonic argon calculated by MCDF. 
  $R_{\rm{\bar{p}}}$ is a measure  for the distance of the antiproton to nucleus 
  (one a.u.=$0.53\cdot 10^{-10}~m$) and $Z^{\rm{\bar{p}}}_{eff}$ the actual nuclear charge screened 
  by the electrons. For example, the first antiprotonic $\Delta n_{\rm{\bar{p}}}=1$ de--excitation 
  step being able to ionise the first 2s electron ($E_B=748~eV$) is the (27--26) transition 
  ($\Delta E_{n_{\rm{\bar{p}}}\rightarrow n_{\rm{\bar{p}}}-1}=850\,eV).$
  Dots (...) denote fully occupied lower--lying electron states. }
\label{tab:pbar_Ar_electrons}
\setlength{\tabcolsep}{2.9mm}
\begin{tabular}{lcccccccc}
\hline\noalign{\smallskip}
                & no. of    & $R_{\rm{\bar{p}}}$ & $Z^{\rm{\bar{p}}}_{eff}(n_{\rm{\bar{p}}})$ & $n_{\rm{\bar{p}}}$ &$\Delta E(Z^{\rm{\bar{p}}}_{eff})$&\multicolumn{3}{c}{$ionisation~energy$ (MCDF)}\\
electronic state& electrons &                    &                          &                    &$n_{\rm{\bar{p}}}\rightarrow n_{\rm{\bar{p}}}-1$& $Z^{e}=18$&$Z^{e}=18+{\rm{\bar{p}}}$&$Z^{e}=17$ \\
                &           &    /\,a.\,u.       &                          &                    &                                                &/\,eV      & /\,eV                   &/\,eV      \\
\hline\noalign{\smallskip}
$1s^22s^22p^63s^23p^6$&18&&&&&16&capture of ${\rm{\bar{p}}}$ &\\
\hline\noalign{\smallskip}
$... 3s^2(3p_{1/2})^2(3p_{3/2})^3$&17&0.3 &15.49&92&{\bf 21} &28&{\bf 15}&14\\
$... 3s^2(3p_{1/2})^2(3p_{3/2})^2$&16&0.2 &16.29&78&{\bf 36} &42&{\bf 26}&25\\
$... 3s^2(3p_{1/2})^2(3p_{3/2})^1$&15&0.16&16.72&69&{\bf 52} &42&{\bf 26}&25\\
$... 3s^2(3p_{1/2})^2$            &14&0.13&16.97&63&{\bf 67} &74&{\bf 57}&53\\
$... 3s^2(3p_{1/2})^1$            &13&0.11&17.23&56&{\bf 96} &90&{\bf 71}&67\\
$... (3s_{1/2})^2$                &12&0.09&17.36&52&{\bf 120}&122&{\bf 100}&95\\
$... (3s_{1/2})^1$                &11&0.07&17.54&46&{\bf 172}&143&{\bf 116}&114\\
\hline\noalign{\smallskip}
$1s^22s^2(2p_{1/2})^2(2p_{3/2})^4$&10&0.040&17.81&35&{\bf 392}&424&{\bf 351}&346\\
$1s^22s^2(2p_{1/2})^2(2p_{3/2})^3$&9 &0.037&17.83&34&{\bf 427}&484&{\bf 405}&400\\
$1s^22s^2(2p_{1/2})^2(2p_{3/2})^2$&8 &0.033&17.87&32&{\bf 515}&547&{\bf 467}&452\\
$1s^22s^2(2p_{1/2})^2(2p_{3/2})^1$&7 &0.030&17.91&30&{\bf 628}&622&{\bf 537}&527\\
$1s^22s^2(2p_{1/2})^2$            &6 &0.029&17.92&29&{\bf 698}&689&{\bf 599}&594\\
$1s^22s^2(2p_{1/2})^1$            &5 &0.027&17.94&28&{\bf 779}&754&{\bf 661}&655\\
$1s^2(2s_{1/2})^2$                &4 &0.025&17.96&27&{\bf 870}&850&{\bf 748}&745\\
$1s^2(2s_{1/2})^1$                &3 &0.023&17.97&26&{\bf 978}&918&{\bf 812}&809\\
\hline\noalign{\smallskip}
$(1s_{1/2}^2)$&2&0.020&18&17&{\bf 3520}&4122&{\bf 3663}&3659\\
$(1s_{1/2}^1)$&1&0.019&18&16&{\bf 4250}&4427&{\bf 3950}&3947\\
\hline\noalign{\smallskip}
\end{tabular}
\end{table*}

In the region of the L~electrons ($n_{\rm{\bar{p}}}\approx 80$) $Z^{\rm{\bar{p}}}_{eff}$ has 
increased to about 15. Around $n_{\rm{\bar{p}}}=80$ the antiprotonic transition energies amount 
to 20--50\,eV, which is now sufficient to remove the M3 electrons (Table\,\ref{tab:pbar_Ar_electrons} 
and Fig.\,\ref{fig:argon_en_pbar_Zeff}). After complete depletion of the M shell, arriving at the ionisation 
state Ar$^{8+}$, the antiproton is situated well inside the L shell. Because of the screening of one 
nuclear charge by the antiproton, the electronic L shell experiences rather the Coulomb field of a 
Cl$^{7+}$ ion. The binding energy for one electron in a full L shell is 350\,eV. 

In the region of the K electrons, $Z^{\rm{\bar{p}}}_{eff}$ for the antiproton exceeds already 17.
At $n_{\rm{\bar{p}}}\cong 35$ the energy of a circular transition reaches 350\,eV  now being 
able to remove L electrons. After removal of the L shell, the argon atom is highly ionized  
($\rm{\bar{p}}$Ar$^{16+}$\,=\,"Cl"$^{15+}$) with the antiproton far inside the K shell 
experiencing the full nuclear charge $Z=18$. The screening of one nuclear charge by the antiproton reduces the 
electron binding energies of Ar$^{16+}$ and Ar$^{17+}$ from 4122 and 4427\,eV for the $1s^{2}$ and $1s^{1}$ 
configurations to 3659 and 3947\,eV, respectively. Removal of the first K electron by a circular transition
is not yet possible for $\rm{\bar{p}}$Ar$^{16+}$(17--16). It requires 4 or more additional electrons 
in the L--shell. A single K electron (H--like "Cl") can be removed first by the $\rm{\bar{p}}$Ar$^{17+}$(16--15) 
transition (4250\,eV).

\subsection{Antiprotonic Cascade in Krypton (Z=36)}
\label{subsec:Kr_cascade}

In krypton, the antiprotonic cascade is assumed to start at about the electronic N\,shell. With 
$n_{\rm{\bar{p}}}\approx 170$ and the effective charge $Z^{\rm{\bar{p}}}_{eff}\approx 25$ circular 
transitions in this region have energies around 7\,eV. The binding energies of 4s and 4p electrons 
are 32 and 15\,eV, respectively (Table\,\ref{tab:Zeff_Kr}). The calculations of electronic and antiprotonic orbital 
energies and radii show that step by step ionisation of the $4p_{3/2}$, $4p_{1/2}$  and $4s$ electrons by 
$\Delta n_{\rm{\bar{p}}}=1$ transitions happens for $Z^{\rm{\bar{p}}}_{eff}\approx 28-34$ corresponding  
to $n_{\rm{\bar{p}}}\approx 150-82$ (Table\,\ref{tab:pbar_Kr_electrons} and Fig.\,\ref{fig:krypton_en_pbar_Zeff}). 
At this stage the antiproton finds itself within the L shell and just about outside the K shell. The 
absence of the screening by the N electrons increases the binding energy of the M\,shell for 
3s, 3p, and 3d electrons from 305, 230, and 102\,eV to 438, 363, and 235\,eV, respectively. With 
the antiproton inside the M\,shell, the effective charge for M\,electrons is lowered by one unit and 
so the binding energies are reduced by about 3.7\%.

\begin{figure} [b]
\resizebox{0.48\textwidth}{!}{\includegraphics{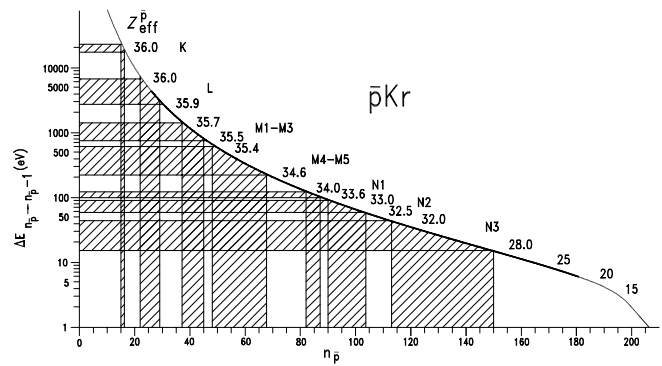}}
\caption[Energies of $\Delta n_{\rm{\bar{p}}}=1$ transitions in antiprotonic krypton] 
 {Energies of $\Delta n_{\rm{\bar{p}}}=1$ transitions in antiprotonic krypton.}
\label{fig:krypton_en_pbar_Zeff}
\end{figure}

\begin{table*}[t]
\caption
 {State and cascade dependent ionisation energies in antiprotonic krypton.}
\label{tab:pbar_Kr_electrons}
\setlength{\tabcolsep}{2.8mm}
\begin{tabular}[t]{lcccccccc}
\hline\noalign{\smallskip}
                & no. of    & $R_{\rm{\bar{p}}}$ & $Z^{\rm{\bar{p}}}_{eff}(n_{\rm{\bar{p}}})$ & $n_{\rm{\bar{p}}}$ &$\Delta E(Z^{\rm{\bar{p}}}_{eff})$              &\multicolumn{3}{c}{$ionisation~energy$ (MCDF)}\\
electronic state& electrons &                    &                          &                    &$n_{\rm{\bar{p}}}\rightarrow n_{\rm{\bar{p}}}-1$& $Z^{e}=36$&$Z^{e}=36+{\rm{\bar{p}}}$&$Z^{e}=35$ \\
                &           &    /\,a.\,u.       &                          &                    &                                                &/\,eV      & /\,eV                   &/\,eV      \\
\hline\noalign{\smallskip}
$... 4s^2(4p_{1/2})^2(4p_{3/2})^4$&36&&&&&13&capture of ${\rm{\bar{p}}}$ &\\
\hline\noalign{\smallskip}
$... 4s^2(4p_{1/2})^2(4p_{3/2})^3$&35&0.46 &28  &150&{\bf 16} &24&{\bf 11}   &11\\
$... 4s^2(4p_{1/2})^2(4p_{3/2})^2$&34&0.31 &30  &130&{\bf 30} &34&{\bf 20}   &20\\
$... 4s^2(4p_{1/2})^2(4p_{3/2})^1$&33&0.23 &32  &113&{\bf 46} &50&{\bf 35}   &34\\
$... 4s^2(4p_{1/2})^2$            &32&0.18 &32.5&104&{\bf 60} &64&{\bf 47}   &46\\
$... 4s^2(4p_{1/2})^1$            &31&0.16 &33.0&90 &{\bf 88} &77&{\bf 61}   &59\\
$... 4s_{1/2})^2$                 &30&0.12 &33.6&87 &{\bf 101}&106&{\bf 86}  &85\\
$... 4s_{1/2})^1$                 &29&0.11 &34.0&82 &{\bf 119}&124&{\bf 103} &102\\
\hline\noalign{\smallskip}
$... (3d_{3/2})^4(3d_{5/2})^6$&28&0.075 &34.68&68&{\bf 211} &231&{\bf 191} &191\\
$... (3d_{3/2})^4(3d_{5/2})^5$&27&0.070 &34.85&64&{\bf 253} &272&{\bf 229} &229\\
$... (3d_{3/2})^4(3d_{5/2})^4$&26&0.060 &34.93&62&{\bf 278} &302&{\bf 257} &257\\
$... (3d_{3/2})^4(3d_{5/2})^3$&25&0.056 &35.05&59&{\bf 323} &348&{\bf 300} &300\\
$... (3d_{3/2})^4(3d_{5/2})^2$&24&0.052 &35.13&57&{\bf 359} &388&{\bf 338} &337\\
$... (3d_{3/2})^4(3d_{5/2})^1$&23&0.046 &35.23&54&{\bf 418} &446&{\bf 395} &393\\

$... (3d_{3/2})^4$            &22&0.043 &35.27&52&{\bf 468} &491&{\bf 436} &435\\
$... (3d_{3/2})^3$            &21&0.040 &35.33&50&{\bf 527} &546&{\bf 489} &488\\
$... (3d_{3/2})^2$            &20&0.038 &35.37&49&{\bf 560} &584&{\bf 525} &524\\
$... (3d_{3/2})^1$            &19&0.037 &35.41&48&{\bf 597} &639&{\bf 578} &577\\
\hline\noalign{\smallskip}
$... 3s^2(3p_{1/2})^2(3p_{3/2})^4$&18&0.031&35.52&45&{\bf 732} &784 &{\bf 716} &715\\
$... 3s^2(3p_{1/2})^2(3p_{3/2})^3$&17&0.029&35.55&44&{\bf 785} &837 &{\bf 767} &766\\
$... 3s^2(3p_{1/2})^2(3p_{3/2})^2$&16&0.028&35.56&43&{\bf 843} &875 &{\bf 804} &803\\
$... 3s^2(3p_{1/2})^2(3p_{3/2})^1$&15&0.027&35.60&42&{\bf 994} &936 &{\bf 862} &861\\
$... 3s^2(3p_{1/2})^2$            &14&0.025&35.63&40&{\bf 1108}&996 &{\bf 919} &918\\
$... 3s^2(3p_{1/2})^1$            &13&0.024&35.66&39&{\bf 1192}&1050&{\bf 970} &969\\
$... (3s_{1/2})^2$                &12&0.023&35.68&38&{\bf 1225}&1150&{\bf 1067}&1066\\
$... (3s_{1/2})^1$                &11&0.022&35.71&37&{\bf 1327}&1204&{\bf 1119}&1118\\
\hline\noalign{\smallskip}
$1s^22s^2(2p_{1/2})^2(2p_{3/2})^4$&10&0.013&35.88&29&{\bf 2790}&2972&{\bf 2732}&2728\\
$1s^22s^2(2p_{1/2})^2(2p_{3/2})^3$&9 &0.012&35.90&28&{\bf 3108}&2924&{\bf 2892}&2887\\
$1s^22s^2(2p_{1/2})^2(2p_{3/2})^2$&8 &0.011&35.92&27&{\bf 3473}&3089&{\bf 3002}&2997\\
$1s^22s^2(2p_{1/2})^2(2p_{3/2})^1$&7 &0.010&35.94&26&{\bf 3902}&3203&{\bf 3202}&3168\\
$1s^22s^2(2p_{1/2})^2$            &6 &0.009&35.96&25&{\bf 4402}&3380&{\bf 3371}&3366\\
$1s^22s^2(2p_{1/2})^1$            &5 &0.009&35.97&24&{\bf 4587}&3590&{\bf 3532}&3527\\
$1s^2(2s_{1/2})^2$                &4 &0.008&35.98&23&{\bf 5683}&3756&{\bf 3733}&3728\\
$1s^2(2s_{1/2})^1$                &3 &0.007&36   &22&{\bf 6585}&3963&{\bf 3875}&3869\\
\hline\noalign{\smallskip}
$(1s_{1/2}^2)$                    &2&0.004&36&16&{\bf 17248}&17309&{\bf 16341}&16315\\
$(1s_{1/2}^1)$                    &1&0.003&36&15&{\bf 21081}&17959&{\bf 16960}&16937\\
\hline\noalign{\smallskip}
\end{tabular}
\end{table*}

The transitions $(n_{\rm{\bar{p}}}\rightarrow n_{\rm{\bar{p}}}-1)$ for $n_{\rm{\bar{p}}}=68$ 
to $n_{\rm{\bar{p}}}=48$ have sufficient energy to remove the M electrons leaving the krypton atom 
in a state $\rm{\bar{p}}$Kr$^{26+}$\,=\,"Br"$^{25+}$. For this ionisation state the L--electron binding 
energies increase substantially by 53\% for 2s and 60\% for 2p compared to the neutral 
atom (Table\,\ref{tab:Zeff_Kr}).

After crossing the electron K shell the antiproton sees almost the full nuclear charge of $Z=36$. 
The antiprotonic transitions from $n_{\rm{\bar{p}}}=29$ to 22 are able to remove the L electrons 
successively. So when the antiproton reaches $n_{\rm{\bar{p}}}=17$ the krypton atom is expected to 
possess only K electrons. The K electrons now have a binding energy of 16341\,eV for the 1s$^{2}$ and 
16960\,eV for the 1s$^{1}$ configurations being 13\% and 17.5\% larger than for neutral krypton. 
Circular transitions starting from $n_{\rm{\bar{p}}}\leq 16$ can now ionise any K electron.

From this analysis, we infer that the $\Delta n = 1$ antiprotonic transitions, having enough energy to 
ionize the N--, M--, L--, and K--shell electrons, start at $n_{\rm{\bar{p}}}\approx$~150, 68, 29, 
and 16, respectively, with a corresponding nuclear charge of $Z^{\rm{\bar{p}}}_{eff}=$~28, 35, 35.9, 
and 36 (Fig.~\ref{fig:krypton_en_pbar_Zeff}).
After complete removal of the M electrons, the antiproton has reached the spatial region of the electronic 
K shell, i.\,e., $n_{\rm{\bar{p}}} \approx 40$ with $Z^{\rm{\bar{p}}}_{eff}\approx 36$. 
This means, that, although the maximum spatial overlap of the antiprotonic
wave function with the K-- and L--shell electrons occurs at
$n{\rm_{\bar{p}}}\approx 42$ and 84, respectively, $\Delta n_{\rm{\bar{p}}}=1$
transitions do not possesss enough energy to ionise the L electrons.
The L--shell ionisation takes place when the antiproton has reached
$n_{\rm{\bar{p}}}\approx 28$ with $Z^{\rm{\bar{p}}}_{eff}\approx 36$. M-- and N--shell 
ionisation show a similar behaviour. The effective charge seen by the M electrons is
obtained from these calculations corrected by nearly one unit to take into account 
the screening by the antiproton's charge (Table\,\ref{tab:Zeff_Kr}).

\subsection{Antiprotonic Cascade in Xenon (Z=54)}
\label{subsec:Xe_cascade}

The antiprotonic cascade in xenon is more complex than that of argon and krypton because of the larger 
number of electrons. Assuming that it starts at about $n_{e}=5$ the effective charge
acting on the 5s and 5p electrons is $Z^{e}_{eff}\approx 20$ and 15, respectively (Table\,\ref{tab:Zeff_Xe}).
The $\Delta n_{\rm{\bar{p}}}=1$ transitions that possess sufficient energy to ionise O--shell electrons 
with binding energies of 12--27\,eV will start at $n_{\rm{\bar{p}}}\cong 180$ where 
$Z^{\rm{\bar{p}}}_{eff}\approx 5$ (Fig.\,\ref{fig:xenon_en_pbar_Zeff} and 
Table\,\ref{tab:pbar_Xe_electrons}). During antiproton capture it is likely that due to 
shake--off effects more than one electron is emitted from the $n_{e}=5$ shell. 
In the absence of only one 5p electron the binding energy of 5s increases by 16\,eV and 
$Z^{\rm{\bar{p}}}_{eff}$ to 16.5. This means that the antiproton has to reach $n_{\rm{\bar{p}}}=87$ 
to remove a 5s electron by a circular transition.

\begin{table*}[t]
\caption
 {State and cascade dependent ionisation energies in antiprotonic xenon.}
\label{tab:pbar_Xe_electrons}
\setlength{\tabcolsep}{1.8mm}
\begin{tabular}[t]{lcccccccc}
\hline\noalign{\smallskip}
                & no. of    & $R_{\rm{\bar{p}}}$ & $Z^{\rm{\bar{p}}}_{eff}(n_{\rm{\bar{p}}})$ & $n_{\rm{\bar{p}}}$ &$\Delta E(Z^{\rm{\bar{p}}}_{eff})$              &\multicolumn{3}{c}{$ionisation~energy$ (MCDF)}\\
electronic state& electrons &                    &                          &                    &$n_{\rm{\bar{p}}}\rightarrow n_{\rm{\bar{p}}}-1$& $Z^{e}=54$&$Z^{e}=54+{\rm{\bar{p}}}$&$Z^{e}=53$ \\
                &           &    /\,a.\,u.       &                          &                    &                                                &/\,eV      & /\,eV                   &/\,eV      \\
\hline\noalign{\smallskip}
$... 5s^2(5p_{1/2})^2(5p_{3/2})^4$&54&&&&&11&capture of ${\rm{\bar{p}}}$ &\\
\hline\noalign{\smallskip}
$... 5s^2(5p_{1/2})^2(5p_{3/2})^{1-3}$&51--53&0.22--0.51     &43--40&144--180&{\bf } &42--21&{\bf 31--11} &29--10\\
$... 5s^2(5p_{1/2})^{1,2}$&49,50&0.19,0.22 &46,45&126,134    &{\bf } &66,54  &{\bf 53,42} &51,40\\
$... (5s_{1/2})^{1-2}$    &47,48&0.14,0.15 &48.2,47.4&109,115&{\bf } &104,90 &{\bf 88,74} &86,72\\
\hline\noalign{\smallskip}
$... (4d_{5/2})^{1-6}$&41--46&0.064--0.10 &51.3--49.8&77--94&{\bf } &314--178&{\bf 284--154} &279--150\\
$... (4d_{3/2})^{1-4}$&37--40&0.051--0.060&51.9--51.4&69--75&{\bf } &434--344&{\bf 399--312} &393--307\\
\hline\noalign{\smallskip}
$... 4s^2(4p_{1/2})^2(4p_{3/2})^4$&36&0.044 &52.27&64&{\bf 631} &550&{\bf 510} &504\\
$... 4s^2(4p_{1/2})^2(4p_{3/2})^3$&35&0.042 &52.33&63&{\bf 653} &587&{\bf 546} &540\\
$... 4s^2(4p_{1/2})^2(4p_{3/2})^2$&34&0.041 &52.39&62&{\bf 678} &611&{\bf 570} &564\\
$... 4s^2(4p_{1/2})^2(4p_{3/2})^1$&33&0.040 &52.45&61&{\bf 706} &650&{\bf 608} &602\\
$... 4s^2(4p_{1/2})^2$ &32&0.037 &52.57 &59 &{\bf 764} &700&{\bf 654} &648\\
$... 4s^2(4p_{1/2})^1$ &31&0.035 &52.63 &58 &{\bf 795} &736&{\bf 689} &683\\
$... 4s_{1/2})^2$      &30&0.033 &52.74 &56 &{\bf 869} &818&{\bf 768} &762\\
$... 4s_{1/2})^1$      &29&0.032 &52.79 &55 &{\bf 912} &856&{\bf 805} &799\\
\hline\noalign{\smallskip}
$... (3d_{3/2})^4(3d_{5/2})^6$&28&0.023 &53.18&47&{\bf 1456} &1494&{\bf 1400} &1397\\
$... (3d_{3/2})^4(3d_{5/2})^5$&27&0.022 &53.20&45&{\bf 1660} &1587&{\bf 1489} &1488\\
$... (3d_{3/2})^4(3d_{5/2})^4$&26&0.021 &53.27&44&{\bf 1780} &1647&{\bf 1548} &1547\\
$... (3d_{3/2})^4(3d_{5/2})^3$&25&0.020 &53.30&43&{\bf 1896} &1742&{\bf 1640} &1639\\
$... (3d_{3/2})^4(3d_{5/2})^2$&24&0.019 &53.35&42&{\bf 2038} &1815&{\bf 1711} &1710\\
$... (3d_{3/2})^4(3d_{5/2})^1$&23&0.018 &53.40&41&{\bf 2195} &1919&{\bf 1813} &1812\\
$... (3d_{3/2})^4$            &22&0.017 &53.43&40&{\bf 2367} &2023&{\bf 1913} &1913\\
$... (3d_{3/2})^3$            &21&0.016 &53.48&39&{\bf 2559} &2127&{\bf 2014} &2014\\
$... (3d_{3/2})^2$            &20&0.015 &53.50&38&{\bf 2770} &2195&{\bf 2080} &2079\\
$... (3d_{3/2})^1$            &19&0.0146&53.55&37&{\bf 2986} &2300&{\bf 2082} &2181\\
\hline\noalign{\smallskip}
$... 3s^2(3p_{1/2})^2(3p_{3/2})^4$&18&0.0140&53.59&36&{\bf 3250} &2556&{\bf 2432} &2431\\
$... 3s^2(3p_{1/2})^2(3p_{3/2})^3$&17&0.0132&53.62&35&{\bf 3547} &2651&{\bf 2525} &2524\\
$... 3s^2(3p_{1/2})^2(3p_{3/2})^2$&16&0.0124&53.66&34&{\bf 3873} &2713&{\bf 2585} &2584\\
$... 3s^2(3p_{1/2})^2(3p_{3/2})^1$&15&0.0117&53.69&33&{\bf 4247} &2811&{\bf 2681} &2680\\
$... 3s^2(3p_{1/2})^2$            &14&0.0110&53.72&32&{\bf 4664} &2975&{\bf 2838} &2836\\
$... 3s^2(3p_{1/2})^1$            &13&0.0100&53.75&31&{\bf 5142} &3068&{\bf 2928} &2926\\
$...(3s_{1/2})^2$                 &12&0.0097&53.78&30&{\bf 5687} &3242&{\bf 3098} &3096\\
$...(3s_{1/2})^1$                 &11&0.0091&53.81&29&{\bf 6413} &3333&{\bf 3186} &3184\\
\hline\noalign{\smallskip}
$1s^22s^2(2p_{1/2})^2(2p_{3/2})^4$&10&0.0080&53.87&27&{\bf 7821} &7660&{\bf 7340}&7338\\
$1s^22s^2(2p_{1/2})^2(2p_{3/2})^3$&9 &0.0070&53.89&26&{\bf 8783} &7931&{\bf 7604}&7602\\
$1s^22s^2(2p_{1/2})^2(2p_{3/2})^2$&8 &0.0068&53.91&25&{\bf 9909} &8103&{\bf 7774}&7772\\
$1s^22s^2(2p_{1/2})^2(2p_{3/2})^1$&7 &0.0063&53.94&24&{\bf 11139}&8382&{\bf 8046}&8044\\
$1s^22s^2(2p_{1/2})^2$            &6 &0.0058&53.96&23&{\bf 12812}&8971&{\bf 8604}&8601\\
$1s^22s^2(2p_{1/2})^1$            &5 &0.0053&53.98&22&{\bf 14726}&9243&{\bf 8869}&8866\\
$1s^2(2s_{1/2})^2$                &4 &0.0049&54   &21&{\bf 16827}&9580&{\bf 9200}&9196\\
$1s^2(2s_{1/2})^1$                &3 &0.0044&54   &20&{\bf 19557}&9810&{\bf 9425}&9421\\
\hline\noalign{\smallskip}
$(1s_{1/2}^2)$                    &2&0.0005&54&16&{\bf 39028}&40269&{\bf 38719}&38715\\
$(1s_{1/2}^1)$                    &1&0.0003&54&15&{\bf 47708}&41300&{\bf 41300}&39722\\
\hline\noalign{\smallskip}
\end{tabular}
\end{table*}

In Fig.\,\ref{fig:xenon_en_pbar_Zeff} the antiprotonic transition energies 
$(n_{\rm{\bar{p}}}\rightarrow n_{\rm{\bar{p}}}-1)$ are given for specific values of
$Z^{\rm{\bar{p}}}_{eff}$. With an empty O shell the binding energy of 4d electrons increases 
to 71\,eV and the antiproton has to descend to $n_{\rm{\bar{p}}}\approx 90$ with 
$Z^{\rm{\bar{p}}}_{eff}\approx 50$ to ionise $4d$ electrons by $\Delta n_{\rm{\bar{p}}}=1$ 
Auger emission. At $n_{\rm{\bar{p}}}\approx 55$ the last N electron may be removed and  
the xenon atom is 26--fold ionised.

\begin{figure*} [t]
\begin{center}
\resizebox{0.75\textwidth}{!}{\includegraphics{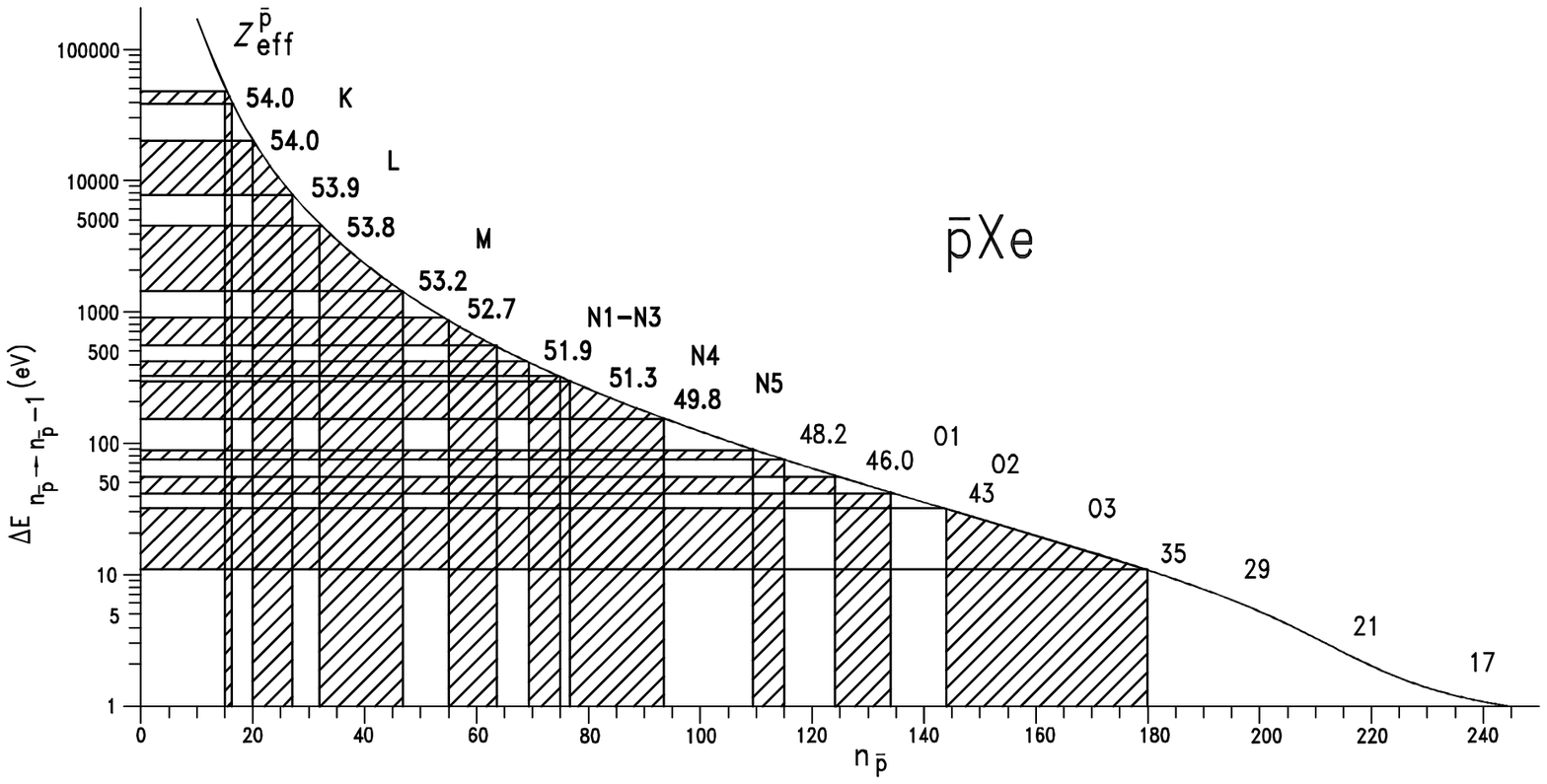}}
\caption[Energies of $\Delta n_{\rm{\bar{p}}}=1$ transitions in antiprotonic xenon] 
 {Energies of $\Delta n_{\rm{\bar{p}}}=1$ transitions in antiprotonic xenon.}
\label{fig:xenon_en_pbar_Zeff}
\end{center}
\end{figure*}

After removal of the N electrons the antiproton has crossed the electronic L and M shell and is 
about to enter the K shell. The binding energy of the M electrons increases to 1413 and 1401\,eV for 
electronic $3d_{3/2}$ and $3d_{5/2}$ states, respectively. Ionisation of the M shell by circular 
transitions starts at $n_{\rm{\bar{p}}}=47$ and $Z^{\rm{\bar{p}}}_{eff}\approx 53$. When reaching 
$n_{\rm{\bar{p}}}\approx 29$ the M shell may be completely ionised. The antiproton, being now far 
inside the K shell, experiences the full nuclear charge independent of the status of the electron shell.

With only L and K shells remaining, the $\rm{\bar{p}}$Xe atom is 44--fold ionised and the electron binding 
energy of the 2p$_{3/2}$, 2p$_{1/2}$, and 2s state increases to 7340, 7647, and 7946\,eV, respectively. 
Auger emission by $\Delta n_{\rm{\bar{p}}}=1$ starts at $n_{\rm{\bar{p}}}=27$, where the antiproton's 
circular transition of 7780\,eV exceeds the electronic binding energy of 7340\,eV 
(see Table\,\ref{tab:pbar_en_dn1}). When the antiproton reaches $n_{\rm{\bar{p}}}=22$ 
the L shell may be completely depleted.  

In the case that such a 52--fold ionised state is reached, the K--electron binding 
energies for the 1s$^{2}$ and the 1s$^{1}$ configurations become 40271 and 41300\,eV.  
Screening of one charge unit by the antiproton reduces these energies by 2.9\% to 
38793 and 39785\,eV. The antiprotonic transition 16--15 having the energy of 39028\,eV 
is able to eject one K electron only if the second one is still present.

\begin{table*}
\caption
 {Electron binding energies $BE$ and effective nuclear charge $Z^e_{eff}$ acting on the electrons of an 
  Ar atom for different electronic configurations without and with ({\it in italic}) an antiproton present 
  in a level $n_{\rm{\bar{p}}}$ as indicated in the last row.}
\label{tab:Zeff_Ar}
\setlength{\tabcolsep}{2.0mm}
\begin{tabular}{lrrrrrrrrrrrr}
\hline\noalign{\smallskip}
levels&$BE$(eV)&$Z^e_{eff}$&$BE$(eV)&$Z^e_{eff}$&$BE$(eV)&$Z^e_{eff}$&$BE$(eV)&$Z^e_{eff}$&$BE$(eV)&$Z^e_{eff}$&$BE$(eV)&$Z^e_{eff}$\\
\hline\noalign{\smallskip}
1s        & 3239 & 17.4 & 3413 & 17.5 & 3747 & 17.5 & 3942 & 17.5 & 4122 & 17.6 & 4427 & 18.0\\
          &      &      & \it 3001 & \it 16.4 & \it 3309 & \it 16.5 &\it 3490 &\it 16.6 &\it 3659 &\it 16.6 &\it 3947 &\it 17.0\\
\hline\noalign{\smallskip}
2s        & 337  & 14.6 & 500  & 13.5 & 724  & 15.9 & 850  & 16.4 & &  & &\\
          &      &      & \it 420 & \it 13.7 & \it 628 & \it 14.9 &\it 745  &\it 15.4 & &  & &\\
2p$_{1/2}$& 262  & 13.4 & 426  & 13.5 & 687  & 15.2 & & & & & &\\
          &      &      & \it 351 & \it 12.5 & \it 595 & \it 14.2 & & & & & &\\
2p$_{3/2}$& 260  & 13.3 & 425  & 13.5 & & & & & & & &\\
          &      &      & \it 349 & \it 12.5 & & & & & & & &\\
\hline\noalign{\smallskip}
3s        & 35   & 9.5  &      &      & & & & & & & &\\
3p$_{1/2}$& 16.2 & 7.5  &      &      & & & & & & & &\\
3p$_{3/2}$& 16   & 7.5  &      &      & & & & & & & &\\
\hline\noalign{\smallskip}
{\small no. of electrons}& \multicolumn{2}{c}{18} & \multicolumn{2}{c}{10} &
                          \multicolumn{2}{c}{6} & \multicolumn{2}{c}{4}  & 
                          \multicolumn{2}{c}{2}  & \multicolumn{2}{c}{1}  \\
{\small configuration}  & \multicolumn{2}{c}{$\rm 1s^2 2s^2 2p^6 3s^23p^6$}    &
                          \multicolumn{2}{c}{$\rm 1s^2 2s^2 2p^6$}    & 
                          \multicolumn{2}{c}{$\rm 1s^2 2s^2 2p^2$} &
                          \multicolumn{2}{c}{$\rm 1s^2 2s^2$} & 
                          \multicolumn{2}{c}{$\rm 1s^2$} &
                          \multicolumn{2}{c}{$\rm 1s^1$}\\[2mm]
\hline\noalign{\smallskip}
$n_{\rm{\bar{p}}}$      & \multicolumn{2}{c}{} &  \multicolumn{2}{c}{\it 15} & 
                          \multicolumn{2}{c}{\it 15} &  \multicolumn{2}{c}{\it 15} & 
                          \multicolumn{2}{c}{\it 15} &  \multicolumn{2}{c}{\it 15}\\
\hline\noalign{\smallskip}
\end{tabular}
\end{table*}

\begin{table*}
\caption[Electron binding energies $BE$ and effective nuclear charge $Z^e_{eff}$ acting on the electrons of Kr] 
 {Electron binding energies $BE$ and effective nuclear charge $Z^e_{eff}$ acting on the electrons of Kr.
  For explanation of symbols see Table\,\ref{tab:Zeff_Ar}.}
\label{tab:Zeff_Kr}
\setlength{\tabcolsep}{2.0mm}
\begin{tabular}{lrrrrrrrrrrrr}
\hline\noalign{\smallskip}
levels&$BE$(eV)&$Z^e_{eff}$&$BE$(eV)&$Z^e_{eff}$&$BE$(eV)&$Z^e_{eff}$&$BE$(eV)&$Z^e_{eff}$&$BE$(eV)&$Z^e_{eff}$&$BE$(eV)&$Z^e_{eff}$\\
\hline\noalign{\smallskip}
1s        & 14415& 35.3 & 14532& 35.3 & 15678& 35.3 & 16922& 35.6 & 17309 & 35.6 & 17949 & 36\\
          &      &      &      &      &\it 14750&\it 34.3 &\it 15953&\it 34.6 &\it 16328 &\it 34.6 &\it 16958 &\it 35\\
\hline\noalign{\smallskip}
2s        & 1961 & 32.2 & 2096 & 32.2 & 3166 & 32.8 & 3965 & 34.6 & & & &\\
          &      &      &      &      &\it 2956 &\it 31.8 &\it 3729 &\it 33.5 & & & &\\
2p$_{1/2}$& 1765 & 31.0 & 1899 & 31.0 & 2982 & 31.7 & & & & & &\\
          &      &      &      &      &\it 2779 &\it 30.7 & & & & & &\\
2p$_{3/2}$& 1711 & 30.8 & 1846 & 30.8 & 2972 & 31.5 & & & & & &\\
          &      &      &      &      &\it 2730 &\it 30.5& & & & & &\\
\hline\noalign{\smallskip}
3s        & 305  & 25.4 & 438  & 25.4 & & & & & & & &\\
3p$_{1/2}$& 234  & 23.3 & 367  & 23.4 & & & & & & & &\\
3p$_{3/2}$& 226  & 23.0 & 359  & 23.2 & & & & & & & &\\
3d$_{3/2}$& 103  & 19.0 & 236  & 19.5 & & & & & & & &\\
3d$_{5/2}$& 101  & 18.9 & 235  & 19.4 & & & & & & & &\\
\hline\noalign{\smallskip}
4s        & 32   & 15.0 & & & & & & & & & &\\
4p$_{1/2}$& 15   & 12.0 & & & & & & & & & &\\
\hline\noalign{\smallskip}
{\small no. of electrons}& \multicolumn{2}{c}{36} & \multicolumn{2}{c}{28} &
                          \multicolumn{2}{c}{10} & \multicolumn{2}{c}{4}  & 
                          \multicolumn{2}{c}{2}  & \multicolumn{2}{c}{1}  \\
{\small configuration}  & \multicolumn{2}{c}{$\rm 1s^2 2s^2 2p^6 3s^2$}    &
                          \multicolumn{2}{c}{$\rm 1s^2 2s^2 2p^6 3s^2$}    & 
                          \multicolumn{2}{c}{$\rm 1s^2 2s^2 2p^6$} &
                          \multicolumn{2}{c}{$\rm 1s^2 2s^2$} & 
                          \multicolumn{2}{c}{$\rm 1s^2$} &
                          \multicolumn{2}{c}{$\rm 1s^1$}\\
                        & \multicolumn{2}{c}{$\rm 3p^6 3d^6 4s^2 4p^6$} &  \multicolumn{2}{c}{$\rm 3p^6 3d^6$} & 
                          \multicolumn{2}{c}{} &  \multicolumn{2}{c}{} & 
                          \multicolumn{2}{c}{} &  \multicolumn{2}{c}{}\\
\hline\noalign{\smallskip}
$n_{\rm{\bar{p}}}$      & \multicolumn{2}{c}{} &  \multicolumn{2}{c}{} & 
                          \multicolumn{2}{c}{\it 30} &  \multicolumn{2}{c}{\it 25} & 
                          \multicolumn{2}{c}{\it 16} &  \multicolumn{2}{c}{\it 16}\\
\hline\noalign{\smallskip}
\end{tabular}
\end{table*}

\begin{table*}
\caption
 {Electron binding energies $BE$ and effective nuclear charge $Z^e_{eff}$ acting on the electrons of Xe.
  For explanation of symbols see Table\,\ref{tab:Zeff_Ar}.}
\label{tab:Zeff_Xe}
\setlength{\tabcolsep}{2.0mm}
\begin{tabular}{lrrrrrrrrrrrr}
\hline\noalign{\smallskip}
levels&$BE$(eV)&$Z^e_{eff}$&$BE$(eV)&$Z^e_{eff}$&$BE$(eV)&$Z^e_{eff}$&$BE$(eV)&$Z^e_{eff}$&$BE$(eV)&$Z^e_{eff}$&$BE$(eV)&$Z^e_{eff}$\\
\hline\noalign{\smallskip}
1s        & 34624& 53.3 & 34740& 53.3 & 35500& 53.3 & 37647& 53.3 & 40271& 53.6 & 41300& 54\\
          &      &      &      &      &{\it 34106} &{\it 52.3}&{\it 36197}&{\it 52.3}&{\it 38796}&{\it 52.6}&{\it 39785}& \it 53\\\hline
2s        & 5590 & 50.0 & 5608 & 50   & 6382 & 50.0 & 8303 & 50.9 & & & &\\
          &      &      &      &      &{\it 6042} &{\it 49.0} &{\it 7946} &{\it 49.9} & & & &\\
2p$_{1/2}$& 5149 & 48.9 & 5259 & 49   & 5999 & 49.0 & 7998 & 49.8 & & & &\\
          &      &      &      &      &{\it 5701} &{\it 47.5} &{\it 7647} &{\it 48.8} & & & &\\
2p$_{3/2}$& 4825 & 48.3 & 4937 & 48   & 5675 & 48.5 & 7661 & 49.5 & & & &\\
          &      &      &      &      &{\it 5403} &{\it 47.5} &{\it 7339} &{\it 48.5} & & & &\\
\hline\noalign{\smallskip}
3s        & 1166 & 43.1 & 1280 & 43   & 1962 & 43.6 & & & & & &\\
          &      &      &      &      &{\it 1849} &{\it 42.6} & & & & & &\\
3p$_{1/2}$& 1022 & 41.1 & 1134 & 41   & 1820 & 42 & & & & & &\\
          &      &      &      &      &{\it 1709} &{\it 40.2} & & & & & &\\
3p$_{3/2}$& 960  & 40.3 & 1071 & 40   & 1754 & 41 & & & & & &\\
          &      &      &      &      &{\it 1650} & & & & & & &\\
3d$_{3/2}$& 707  & 37.4 & 819  & 37   & 1511 & 38.4 & & & & & &\\
          &      &      &      &      &{\it 1413} &{\it 37.4} & & & & & &\\
3d$_{5/2}$& 694  & 37.2 & 806  & 37   & 1497 & 38.3 & & & & & &\\
          &      &      &      &      &{\it 1401} &{\it 37.2} & & & & & &\\
\hline\noalign{\smallskip}
4s        & 229  & 33.0 & 338  & 33   & & & & & & & &\\
4p$_{1/2}$& 175  & 30.4 & 284  & 30.5 & & & & & & & &\\
4p$_{3/2}$& 163  & 29.6 & 271  & 30   & & & & & & & &\\
4d$_{3/2}$& 73.7 & 24.1 & 187  & 24   & & & & & & & &\\
4d$_{5/2}$& 71.7 & 23.8 & 80   & 24   & & & & & & & &\\
\hline\noalign{\smallskip}
5s        & 27.5 & 19.6 &      &      & & & & & & & &\\
5p$_{1/2}$& 13.4 & 16.2 &      &      & & & & & & & &\\
5p$_{3/2}$& 12.0 & 15.5 &      &      & & & & & & & &\\
\hline\noalign{\smallskip}
{\small no. of electrons}& \multicolumn{2}{c}{54} & \multicolumn{2}{c}{46} &
                          \multicolumn{2}{c}{28} & \multicolumn{2}{c}{10}  & 
                          \multicolumn{2}{c}{2}  & \multicolumn{2}{c}{1}  \\
{\small configuration}  & \multicolumn{2}{c}{$\rm 1s^2 2s^2 2p^6 3s^2$}    &
                          \multicolumn{2}{c}{$\rm 1s^2 2s^2 2p^6 3s^2$}    & 
                          \multicolumn{2}{c}{$\rm 1s^2 2s^2 2p^6 3s^2$} &
                          \multicolumn{2}{c}{$\rm 1s^2 2s^2 2p^6$} & 
                          \multicolumn{2}{c}{$\rm 1s^2$} &
                          \multicolumn{2}{c}{$\rm 1s^1$}\\
                        & \multicolumn{2}{c}{$\rm 3p^6 3d^{10}4s^2 4p^6$} &  \multicolumn{2}{c}{$\rm 3p^6 3d^{10}4s^2 4p^6$} & 
                          \multicolumn{2}{c}{$\rm 3p^6 3d^{10}$} &  \multicolumn{2}{c}{} & 
                          \multicolumn{2}{c}{} &  \multicolumn{2}{c}{}\\
                        & \multicolumn{2}{c}{$\rm 4d^{10}5s^2 5p^6$} & \multicolumn{2}{c}{$\rm 4d^{10}$} & 
                          \multicolumn{2}{c}{} &  \multicolumn{2}{c}{} & 
                          \multicolumn{2}{c}{} &  \multicolumn{2}{c}{}\\
\hline\noalign{\smallskip}
$n_{\rm{\bar{p}}}$      & \multicolumn{2}{c}{} &  \multicolumn{2}{c}{} & 
                          \multicolumn{2}{c}{\it 30} &  \multicolumn{2}{c}{\it 30} & 
                          \multicolumn{2}{c}{\it 17} &  \multicolumn{2}{c}{\it 16}\\
\hline\noalign{\smallskip}
\end{tabular}
\end{table*}

\section{Discussion of the spectra}
\label{sec:discussion}

\subsection{Antiprotonic Argon}
\label{subsec:discussion_Ar} 

The antiprotonic argon X--ray spectrum (Fig.\,\ref{fig:Ar_Kr_Xe_spec_full}) is dominated by 
saturated circular transitions. The decrease in intensity 
below 3\,keV and towards high energies is due to the detection efficiency (Fig.\,\ref{fig:pbar_Ne_rel_eff}). 
The gap between the strong transitions (18--17) and (15--14) is attributed to K electron emission 
(Fig.\,\ref{fig:Ar_K_fit}). Such a gap is not observable in $\rm{\bar{p}}$Ne because it is below the detection 
threshold. As discussed in Sec.\,\ref{subsec:Ar_cascade} for a circular cascade, the antiproton is able 
to ionise the argon atom up to the state where only the K electrons remain when it reaches 
$n_{\rm{\bar{p}}}\cong 25$ (Table\,\ref{tab:pbar_Ar_electrons}). 

The intensity of the $\rm{\bar{p}}$Ar(16--15) line suggests about 25\% completely ionised argon 
for states with $n_{\rm{\bar{p}}}> 17$ (Table\,\ref{tab:pbar_Ar_lines}). Such a high ionisation 
requires $\Delta n_{\rm{\bar{p}}}\geq 2$ transitions in the upper part of the cascade, but with 
not too small angular momenta, because otherwise $\Delta n_{\rm{\bar{p}}}\gg 1$ radiative de--excitation 
dominates over Auger emission.

\begin{figure}[b]
\resizebox{0.45\textwidth}{!}{\includegraphics{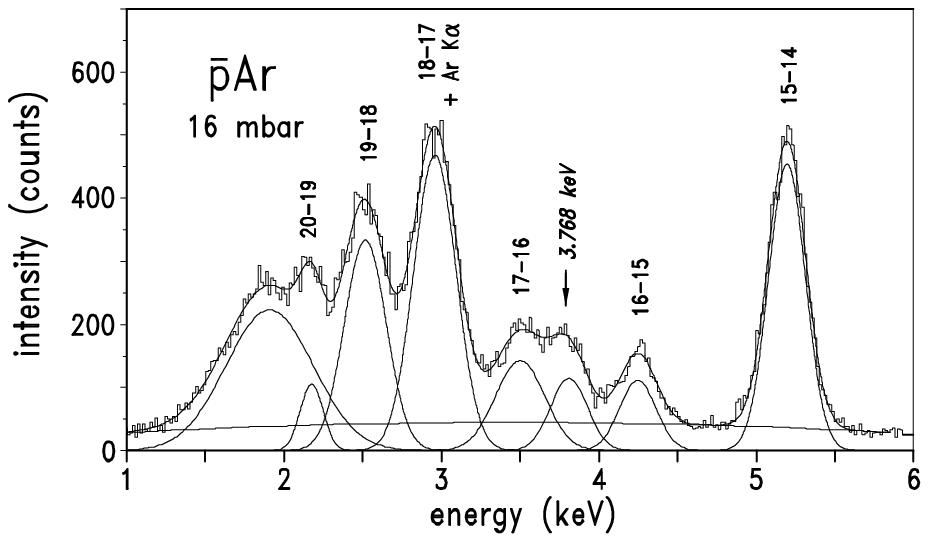}}
  \caption{Spectrum of antiprotonic argon in the region of electronic K X--rays.}  
\label{fig:Ar_K_fit}
\end{figure}

The $\rm{\bar{p}}$Ar(17--16) transition energy strongly depends on the ionisation state. As 
can be seen from Table\,\ref{tab:Zeff_Ar}, the $\rm{\bar{p}}$Ar(17--16) transition is not 
able to ionise a K electron in the case that only 3 or less electrons are present. 
On the other hand any L electron would be ejected by preceeding circular transitions. 
However, inner transitions ($\Delta n_{\rm{\bar{p}}}> 1,\ell\ll\ell_{max}$) are 
able to reach the $n_{\rm{\bar{p}}}=17$ state whith L electrons still present. Ejection 
of the second K electron could proceed via the $\rm{\bar{p}}$Ar(15--14) line, which indeed is 
weakened by 10--20\% as compared to the following transitions. From the $\rm{\bar{p}}$Ar(14--13) 
line onwards the transitions are consistent with a pure radiative cascade without any 
Auger component and the $\rm{\bar{p}}$Ar atom exists as a pure \linebreak
hydrogen--like system.

\begin{table}[h]
\caption{Fluorescence X--rays from argon, krypton, and xenon and the corresponding $Z-1$ atoms.
         The (experimental) energies are taken from\,\cite{des03}.}
\label{tab:fluorescence}
\setlength{\tabcolsep}{1.6mm}
\begin{tabular}[t]{ccccccc}
\hline\noalign{\smallskip}
   & K$\alpha_{1}$ &K$\alpha_{2}$ &K$\beta_{1}$ &K$\beta_{3}$ &L$\alpha_{1}$ &L$\beta_{3}$ \\
\hline\noalign{\smallskip}
Cl &  2622.4       &  2620.9      &   2815.6    & & &\\
Ar &  2957.7       &  2955.6      &   3190.5    & & &\\
\hline\noalign{\smallskip}
Br & 11924.4       & 11877.8      &  13291.6    & 13284.7     & 1480.5        & \\
Kr & 12648.0       & 12595.4      &  14112.8    & 14105.0     & 1585.4        & \\
\hline\noalign{\smallskip}
I  & 28612.3       & 28317.5      &  32295.1    & 32239.7     & 3937.7        & 4120.5\\
Xe & 29778.8       & 29458.3      &  33624.2    & 33563.2     & 4110.1        & 4512.0\\
\hline\noalign{\smallskip}
\end{tabular}
\end{table}

\begin{table}[h]
\caption{Line assignment in the spectrum of antiprotonic argon (Fig.\,\ref{fig:Ar_Kr_Xe_spec_full} and 
         \ref{fig:Ar_K_fit}). The errors are statistical only.}
\label{tab:pbar_Ar_lines}
\setlength{\tabcolsep}{0.6mm}
\begin{tabular}[t]{rclcrclclcc}
\hline\noalign{\smallskip}
\multicolumn{3}{c}{experimental } && \multicolumn{3}{c}{relative} && explanation              && theoretical\\
\multicolumn{3}{c}{energy}        && \multicolumn{3}{c}{intensity}&&                          && energy\\
\multicolumn{3}{c}{(eV)}          && \multicolumn{3}{c}{(\%)}     &&                          && (keV)  \\
\hline\noalign{\smallskip}
2182   &$\pm$& 11                 &&  91.6 &$\pm$ & 14.0          && $\rm{\bar{p}Ar}$(20--19)1s$^{2}$&& 2129\\
2516   &$\pm$&  5                 && 146.9 &$\pm$ & 11.1          && $\rm{\bar{p}Ar}$(19--18)1s$^{2}$&& 2495\\
\hline\noalign{\smallskip}
2949   &$\pm$&  3                 && 166.2 &$\pm$ &  9.5          && $\rm{\bar{p}Ar}$(18--17)/Ar\,K$\alpha$&& 2952/2957\\
\hline\noalign{\smallskip}
3494   &$\pm$& 11                 &&  47.9 &$\pm$ &  3.2          && $\rm{\bar{p}Ar}$(17--16)&& 3523\\
3768   &$\pm$& 10                 &&  36.3 &$\pm$ &  2.4          && \it not identified      && \\
4242   &$\pm$&  5                 &&  27.1 &$\pm$ &  1.0          && $\rm{\bar{p}Ar}$(16--15)&& 4251\\
5197   &$\pm$&  2                 &&  81.8 &$\pm$ &  2.2          && $\rm{\bar{p}Ar}$(15--14)&& 5196\\
6440   &$\pm$&  1                 &&  88.1 &$\pm$ &  2.1          && $\rm{\bar{p}Ar}$(14--13)&& 6441\\
7219   &$\pm$&  38                &&   1.0 &$\pm$ &  0.3          && $\rm{\bar{p}O}$(8--7)   && 7206\\
8117   &$\pm$&  1                 &&  91.7 &$\pm$ &  2.2          && $\rm{\bar{p}Ar}$(13--12)&& 8120\\
10443  &$\pm$&  1                 &&  \multicolumn{3}{c}{100}     && $\rm{\bar{p}Ar}$(12--11)&& 10438\\
11104  &$\pm$&  16                &&   5.5 &$\pm$ &  0.3          && $\rm{\bar{p}O}$(7--6)   && 11110\\
13723  &$\pm$&  2                 &&  97.7 &$\pm$ &  2.1          && $\rm{\bar{p}Ar}$(11--10)&& 13729\\
18560  &$\pm$&  2                 &&  90.3 &$\pm$ &  1.9          && $\rm{\bar{p}Ar}$(10--9) && 18566\\
22887  &$\pm$& 24                 &&   5.0 &$\pm$ &  0.3          && \it not identified      && \\
25977  &$\pm$&  3                 &&  86.6 &$\pm$ &  1.8          && $\rm{\bar{p}Ar}$(9--8)  && 25972\\
\hline\noalign{\smallskip}
\end{tabular}
\end{table}

Because of its high yield the line at the $\rm{\bar{p}}$Ar(18--17) position is interpreted as a 
superposition of argon K$\alpha$ fluorescence radiation and the antiprotonic line (Table\,\ref{tab:fluorescence}).
The energy of the electronic X--rays is consistent with low charge states only. As no K$\beta$ component has 
been found, a significant direct ionisation by the beam is excluded. The high yield of the transition at 
the energy of the $\rm{\bar{p}}$Ar(19--18) remains unexplained. A possible reason maybe an improper treatment of the 
background. 

At 3.77\,keV an additional transition is detected, which appears with an 
intensity comparable to the $\rm{\bar{p}}$Ar(17--16) transition. As can be seen from Table\,\ref{tab:Zeff_Ar} 
it cannot be attributed to an electronic 2p--1s transition for any ionisation state 
or to a $\rm{\bar{p}}$Ar atom. Lines of low intensity were found at 7.2, 11.1, and 22.8\,keV. 
The energies of the first two lines coincide with $\rm{\bar{p}}$O(8--7) and $\rm{\bar{p}}$O(7--6) 
transitions originating from a small water contamination, whereas the third one is not 
identified. No $\Delta n_{\rm{\bar{p}}}=2$ or 3 transitions could be identified in $\rm{\bar{p}}$Ar.

\subsection{Antiprotonic Krypton}
\label{subsec:discussion_Kr}

The high energy part of the krypton spectrum shows a structure similar to that of argon 
(Fig.\,\ref{fig:Ar_Kr_Xe_spec_full}). The impossibility for the antiproton to eject K electrons by a 
circular transition at $n_{\rm{\bar{p}}} > 16$ is evident from the saturated $\rm{\bar{p}}$Kr lines 
(23--22) to (17--16) followed by X--ray suppression for the (16--15) and (15--14) transitions.
Due to the larger energies higher--lying antiprotonic transitions up to at least (25--24) become now 
visible. The intensity reduction for $n_{\rm{\bar{p}}} \ge 26$ is due to the depletion of the L shell. 

Additional intensity is observed in the energy range between 12 and 14\,keV (Fig.\,\ref{fig:Kr_K_xrays_elec}). 
These lines are attributed in most of the cases to electronic X--rays (see below).

\subsubsection{Antiprotonic transitions}

The incomplete suppression of the $\rm{\bar{p}}$Kr(16--15) and \linebreak 
{$\rm{\bar{p}}$Kr(15--14)} lines shows again that the cascade does not proceed exclusively through circular 
states allowing already a partial depletion of the K shell before the antiproton reaches the level 
$n_{\rm{\bar{p}}}=16$ by $\Delta n > 1$ transitions. By the time the antiproton has 
reached $n_{\rm{\bar{p}}} = 14$, however, the observed maximal line yield requires complete 
ionisation (Table\,\ref{tab:pbar_Kr_lines}).

Suppression of X--rays for transitions starting from $n_{\rm{\bar{p}}}>24$ in combination with 
saturated yields of the lines (23--22) to (17--16) suggests that L electron removal takes place 
the same way as for K electrons, however, with about two electrons already removed when reaching the L--emission 
threshold for $\Delta n_{\rm{\bar{p}}}=1$ transitions. Again, such a removal must be caused by 
$\Delta n_{\rm{\bar{p}}}>1$ de--excitation in the higher cascade. Otherwise, as the calculations of electronic 
binding and antiprotonic transition energies show, for a fully occupied L~shell the removal of L electrons by 
Auger emission can start the earliest at $n_{\rm{\bar{p}}}\approx 29$ for circular transitions and the 
lines (24--23), (23--22), and (22--21) would be still suppressed (Table\,\ref{tab:pbar_Kr_electrons}). 
The $\rm{\bar{p}}$Kr(23--22) transition reaches almost the maximum yield, i.\,e., the last L electron is removed 
in the step (24--23).

Below 3.5\,keV down to 1.6\,keV indications of more antiprotonic transitions are seen 
(Fig.\,\ref{fig:Kr_L_xrays_elec}). Such lines can be tentatively assigned to radiative de--excitation 
after complete removal of all M and N electrons, but with the few L electrons present which are bound 
stronger than 3.5\,keV. Interpreting the lines at 1.61 and 2.04\,keV as the $\rm{\bar{p}}$Kr(34--33) and 
$\rm{\bar{p}}$Kr(32--31) transitions, the line yield can be as high as 40\% and 70\%, respectively. Needless 
to say, that a clear distinction between background and X--ray transition is difficult because of the nearby 
detection threshold and the non--resolution of other possible lines.

\subsubsection{K--X--ray region}

\begin{figure} [t]
\resizebox{0.45\textwidth}{!}{\includegraphics{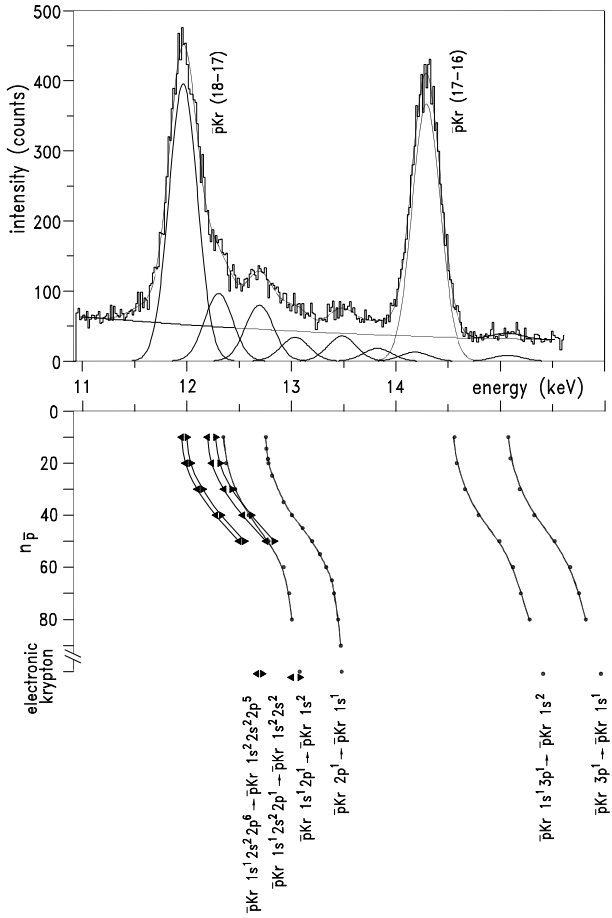}}
\caption{Top -- Spectrum antiprotonic kryton expanded in the energy range of the electronic K X--rays.
         Bottom -- Calculated energies of electronic transitions as a function of the antiproton orbit. 
         Double lines represent the K$\alpha_{1,2}$ doublet. Symbols close to the legend correspond to 
         krypton atoms without an antiproton present.}
\label{fig:Kr_K_xrays_elec}
\end{figure}

The complex structure visible in the range 12 to 14 keV (Fig.\,\ref{fig:Kr_K_xrays_elec}\,--\,top) is mostly 
attributed to electronic X--ray emission during the antiprotonic cascade. For the discussion K transition energies 
of some electron shell configurations are plotted in the lower part of Fig.\,\ref{fig:Kr_K_xrays_elec} as a 
function of the antiprotonic state $n_{\rm{\bar{p}}}$. The decrease in energy of about 0.75\,keV reflects 
the reduction of the effective charge seen by the electrons by one unit from krypton ($Z=36$) to a 
bromine--like ($Z=35$) nucleus when the antiproton penetrates below the rms radii of the K shell. 

The most prominent satellite peak has an energy of 12.287\,keV with a relative intensity of 24\% 
of the \linebreak 
$\rm{\bar{p}}$Kr(18--17) transition (Table\,\ref{tab:pbar_Kr_lines}). Considering first the case of the 
$\rm{1s^1 2p^1} \to \rm{1s^2}$ transition energy in the presence of a complete L shell, the energy can be 
reached only if the antiproton is still at or above the K shell, i.\,e., $n_{\rm{\bar{p}}}\geq 40$, because 
of screening. K hole production at that level requires $\Delta n_{\rm{\bar{p}}}\gg 1$ 
Auger transitions (Table\,\ref{tab:pbar_en_dn1}) and is very unlikely in the presence of L electrons. 
For already significantly ionised krypton with only very few L electrons left, the K$\alpha$ transition energy  
corresponds to an antiproton state at $n_{\rm{\bar{p}}}\approx 20$. Indeed from $n_{\rm{\bar{p}}}\approx 20$ on, 
the $\Delta n_{\rm{\bar{p}}}=2$ Auger transition is able to ionise the K shell. As can be seen from 
Fig.\,\ref{fig:Kr_K_xrays_elec}, no other configuration is able to reach 12.29\,keV. Furthermore, an 
indication of the K$\beta$ analogue is found at 15.09\,keV.

The line found at an energy of 12.696\,keV could be due to K$\alpha$ fluorescence by radiation excitation. 
However, an inclusion of the K$\beta$ line at 14.1\,keV with a K$\alpha$/K$\beta$ intensity 
ratio of 5.8\,\cite{sal74} improves only slightly the fit of the low--energy side of the 
$\rm{\bar{p}}$Kr(17--16) transition. In addition, the energy is about 60\,eV too high for normal krypton 
fluorescence. As discussed earlier\,\cite{sim94} a second Auger emission creating a double K hole before 
radiative de--excitation of one remaining L electron, again at $n_{\rm{\bar{p}}}\approx 20$, leads to such a 
high electron transition energy. This requires a high degree of ionisation at that stage, which is not 
excluded because of the high yield of the $\rm{\bar{p}}$Kr(25--24) and subsequent lines. Because the transition 
occurs in an atom with a double K hole it is an analogue to the hypersatellite case.

If interpreted as a $\rm{\bar{p}}$Kr electronic transition, the line at 12.954\,keV requires almost complete 
depletion at $n_{\rm{\bar{p}}}\approx 40$. In addition, again the production of a double K hole is 
needed to allow a $\rm{2p} \to \rm{1s}$ transition.  Such a high degree of ionisation, however, is 
very unlikely at this stage of the antiprotonic cascade. So, no interpretation is given. 

Capture from the continuum or $\Delta n_{\rm{\bar{p}}}\gg 1$ steps from very high states to about 
$n_{\rm{\bar{p}}}=40$, where screening of one nuclear charge is not yet complete, produces only very low  
ionisation states. In the case of hardly depleted electron shells a K hole leads to K--X--ray emission in 
a bromine--like atom. An energy of 13.5\,keV corresponds to K$\beta$ emission from such systems. The measured 
energy is 13.487\,keV with a relative yield of 8\%. The corresponding K$\alpha$ coincides exactly with the 
$\rm{\bar{p}}$Kr(18--17) transition (relative yield 100\%). However, assuming K$\alpha$/K$\beta$ ratios 
comparable to normal atoms, the K$\alpha$ contribution should be visible as an outstanding high yield 
of the (18--17) line. This is not observed.

\begin{figure} [b]
\resizebox{0.45\textwidth}{!}{\includegraphics{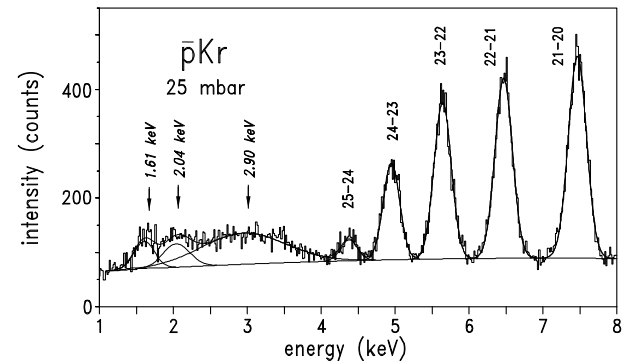}}
\caption{Spectrum of antiprotonic krypton expanded in the energy range of the electronic L X--rays.}
\label{fig:Kr_L_xrays_elec}
\end{figure}

Close to the line at 13.487\,keV is the antiprotonic transition $\rm{\bar{p}}$Al(9--8) having a 
calculated energy of 13.38\,keV. Indeed at 19.5 and 30.1\,keV more $\rm{\bar{p}}$Al transitions are observed. 
But for the line under discussion, first the energy is too high by 100\,eV and secondly the measured 
yield is too high relative to the other $\rm{\bar{p}}$Al transitions. 

Furthermore there is evidence for a line at 13.81\,keV, the energy of which fits to the $\rm{\bar{p}}$Kr(22--20) 
transition \linebreak
(13.869\,keV) in the presence of two K electrons. It is noteworthy that no evidence for any further 
$\Delta n_{\rm{\bar{p}}}=2$ transition was found. A possible $\rm{\bar{p}}$Kr(23--21) transition is hidden 
below $\rm{\bar{p}}$Kr(17--16) line.

A second region with additional low--intensity lines is observed between the $\rm{\bar{p}}$Kr(19--18) and 
$\rm{\bar{p}}$Kr(18--17) transitions (Fig.\,\ref{fig:Ar_Kr_Xe_spec_full}). The largest component amounts to 
$6.1\pm 1.1$\% at $10.641\pm 32$\,keV. 
The energy range fits to electronic X--rays from few--electrons systems having a nuclear charge $Z$ 
in the range of 31--34. The production of such atoms with $Z=35-31$ by antiproton--nucleon annihilation 
at the nuclear surface has been observed by detection of $\gamma$--rays after capture\,\cite{mos86}. 
However, at the time of annihilation the Kr atom can be assumed to be completely ionised as are then the 
daughter nuclei and, therefore, electron refilling has to occur. At the low pressure of 20\,mbar collision rates 
are as low as $\approx 10^{6}/s$. In addition, the cross section for radiative recombination is calculated 
to be negligible\,\cite{bak60}. Hence, no emission of K and L atomic X--rays is expected and an assignment
for the small contaminations in this energy range remains open.

\subsubsection{L--X--ray region}

Below 4\,keV a broad structure is observed together with the indications for higher antiprotonic 
transitions (Fig.\,\ref{fig:Kr_L_xrays_elec}). The energy range fits to electronic L~X--rays at 
various higher charge states (around 2\,keV), which requires the presence of some M electrons. 
A rough estimate for the yield of the L--X--ray intensity is given to be $2\,\pm\,1$ per 
antiprotonic krypton atom.

\begin{table}[t]
\caption{Line assignment in the spectrum of antiprotonic krypton.}
\label{tab:pbar_Kr_lines}
\setlength{\tabcolsep}{0.2mm}
\begin{tabular}[t]{rclcrclclcc}
\hline\noalign{\smallskip}
\multicolumn{3}{c}{experimental } && \multicolumn{3}{c}{relative} && explanation              && theoretical\\
\multicolumn{3}{c}{energy}        && \multicolumn{3}{c}{intensity}&&                          && energy\\
\multicolumn{3}{c}{(eV)}          && \multicolumn{3}{c}{(\%)}     &&                          && (keV)  \\
\hline\noalign{\smallskip}
4369   &$\pm$& 13                 &&  17.6 &$\pm$ &  2.4          && $\rm{\bar{p}Kr}$(25--24)1s$^{2}$&& 4341\\
4939   &$\pm$&  3                 &&  64.2 &$\pm$ &  4.2          && $\rm{\bar{p}Kr}$(24--23)1s$^{2}$&& 4923\\
5627   &$\pm$&  2                 &&  88.9 &$\pm$ &  2.8          && $\rm{\bar{p}Kr}$(23--22)1s$^{2}$&& 5613\\
6447   &$\pm$&  2                 &&  87.1 &$\pm$ &  2.6          && $\rm{\bar{p}Kr}$(22--21)1s$^{2}$&& 6438\\
7445   &$\pm$&  5                 &&  89.7 &$\pm$ &  3.8          && $\rm{\bar{p}Kr}$(21--20)1s$^{2}$&& 7431\\
8645   &$\pm$&  2                 &&  85.0 &$\pm$ &  2.4          && $\rm{\bar{p}Kr}$(20--19)1s$^{2}$&& 8639\\
10133  &$\pm$&  2                 &&  96.2 &$\pm$ &  2.6          && $\rm{\bar{p}Kr}$(19--18)1s$^{2}$&& 10123\\
10642  &$\pm$&  32                &&   6.1 &$\pm$ &  1.1          && \it not identified       &&\\
11968  &$\pm$&  7                 &&  \multicolumn{3}{c}{100}     && $\rm{\bar{p}Kr}$(18--17)1s$^{2}$+"Br"\,K$\alpha$?&& 11967\\
\hline\noalign{\smallskip}
12287  &$\pm$& 33                 &&  24.3 &$\pm$ & 1.5           && $\rm{\bar{p}Kr}$ $n \approx 20$ $\rm{1s2p} \to \rm{1s^2}$ el. && 12290\\
12696  &$\pm$& 24                 &&  20.4 &$\pm$ & 1.4           && $\rm{\bar{p}Kr}$ $n \approx 20$ $\rm{2p} \to \rm{1s}$ el.     && 12760\\
       &     &                    &&       &      &               && K$\alpha$ \textit{fluorescence?} && 12648/12597\\
12954  &$\pm$& 32                 &&   8.6 &$\pm$ & 1.2           && \textit{not identified}          && \\
13487  &$\pm$& 20                 &&   9.0 &$\pm$ & 1.4           && "Br" K$\beta$?                   && \\
13814  &$\pm$& 39                 &&   4.6 &$\pm$ & 1.0           && $\rm{\bar{p}Kr}$(22--20) with $\rm{1s^2}$ el. && 13860\\

14288  &$\pm$&  2                 &&  95.5 &$\pm$ & 2.3           && $\rm{\bar{p}Kr}$(17--16)1s$^{2}$&& 14286\\
\hline\noalign{\smallskip}
15090  &$\pm$& 43                 &&   2.4 &$\pm$ &  0.7          && $\rm{\bar{p}Kr}$ $n \le 14$ $\rm{3p} \to \rm{1s}$ el.      && 15080\\
17261  &$\pm$& 12                 &&  11.6 &$\pm$ &  1.0          && $\rm{\bar{p}Kr}$(16--15)&& 17250\\
19537  &$\pm$& 27                 &&   3.2 &$\pm$ &  0.8          && $\rm{\bar{p}Al}$(8--7)  && 19523\\
21090  &$\pm$&  4                 &&  33.6 &$\pm$ &  1.2          && $\rm{\bar{p}Kr}$(15--14)&& 21084\\
26147  &$\pm$&  3                 &&  89.4 &$\pm$ &  2.3          && $\rm{\bar{p}Kr}$(14--13)&& 26145\\
29052  &$\pm$& 19                 &&   4.4 &$\pm$ &  0.6          && \it not identified      && \\
30146  &$\pm$& 25                 &&   3.4 &$\pm$ &  0.5          && $\rm{\bar{p}Al}$(7--6)  && 30107\\
32962  &$\pm$&  3                 &&  80.5 &$\pm$ &  2.2          && $\rm{\bar{p}Kr}$(13--12)&& 32965\\
\hline\noalign{\smallskip}
\end{tabular}
\end{table}

\subsection{Antiprotonic Xenon}
\label{subsec:discussion_Xe}

In contrast to argon and krypton, the most prominent features of the antiprotonic xenon spectrum 
are two broad structures at 4.5 and 29\,keV (Fig.\,\ref{fig:Ar_Kr_Xe_spec_full}).  The energy range 
corresponds to electronic K and L X--rays of iodine to xenon. The large width of the distribution 
of about 3 and 2\,keV, respectively, indicates that a large variety of ionisation states contributes.  
Antiprotonic transitions are observed as high as (29--28). The low--lying lines having energies 
above 60\,keV exceed the detection range.

\begin{figure} [b]
\resizebox{0.45\textwidth}{!}{\includegraphics{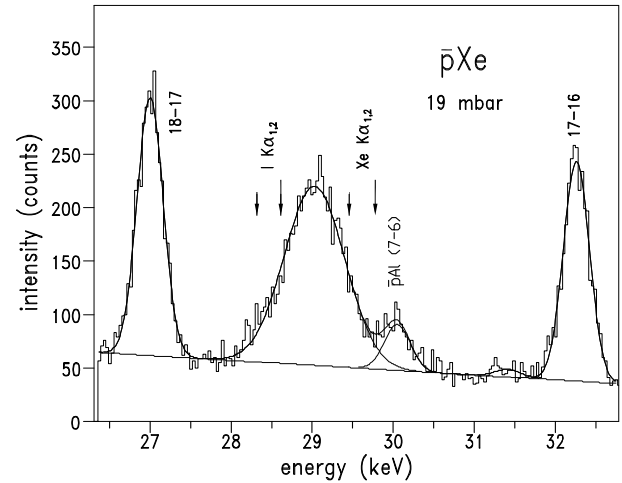}}
\caption{Spectrum of antiprotonic xenon expanded in the energy range of electronic K X--rays. 
         The arrows indicate the energies of the iodine and xenon K$\alpha$ doublets 
         (see Table\,\ref{tab:fluorescence}).}
\label{fig:Xe_Kxrays}
\end{figure}

\begin{figure} [h]
\resizebox{0.45\textwidth}{!}{\includegraphics{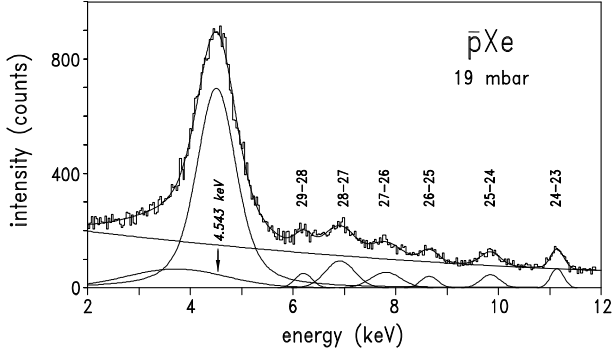}}
\caption{Spectrum of antiprotonic xenon expanded in the energy range of electronic L X--rays.}
\label{fig:Xe_Lxrays}
\end{figure}

\subsubsection{Antiprotonic transitions}

Following the discussion of the antiprotonic cascade in Sec.\,\ref{subsec:Xe_cascade} ionisation of the 
fully occupied M shell can commence at $n_{\rm{\bar{p}}}\approx 68$ and ends with the $\rm{\bar{p}}$Xe(37--36) 
transition. Line yields around 20\% for the transitions $\rm{\bar{p}}$Xe(29--28) to $\rm{\bar{p}}$Xe(24--23) 
constitutes evidence that indeed in some cases the M shell is depleted, when the antiproton reaches 
$n_{\rm{\bar{p}}}=29$ (Table\,\ref{tab:pbar_Xe_electrons}). From $n_{\rm{\bar{p}}}=29$ on also L--shell 
depletion by $\Delta n_{\rm{\bar{p}}}=1$ transitions becomes possible. 

We may note, that the occurance of circular transitions from $n_{\rm{\bar{p}}}=29-27$ requires 
M--shell depletion, whereas the finite yield of the lines $\rm{\bar{p}}$Xe(26--25) to $\rm{\bar{p}}$Xe(17--16) 
is due to the increase of radiative de--excitation with increasing energy. Even more, the relative yields 
coincide rather accurately with the energy dependence of $\Gamma_{X}/\Gamma_{tot}$ as expected 
from Ferell's formula (Fig.\,\ref{fig:pbar_fig04_GA_GX_Gtot}). About two--thirds of the L electrons 
can be removed until the level $n_{\rm{\bar{p}}}=16$ is reached. 

Even assuming atoms with a significant remaining N--shell electron population, the binding energies 
of the K electrons never become small enough to allow Auger emission at $n_{\rm{\bar{p}}}=17$ 
(Table\,\ref{tab:Zeff_Xe}). Therefore, the first de--excitation step being able to emit a K electron 
is the $\rm{\bar{p}}$Xe(16--15) transition. As then expected, the line yield drops to about the fraction 
suggested by $\Gamma_{X}/\Gamma_{tot}$. The observed non--saturation of the $\rm{\bar{p}}$Xe(14--13) yield 
is consistent with incomplete L--shell ionisation at $n_{\rm{\bar{p}}}=16$ as discussed above.  

An approximately linear increase in intensity of the transitions with $n_{\rm{\bar{p}}}\le 15$, as suggested 
by Ferell's formula is not observed. Because of the increasing electron depletion even higher X--ray 
yields are expected. No explanation was found for such a behaviour.

\begin{table}[b]
\caption{Line assignment in the spectrum of antiprotonic xenon.}
\label{tab:pbar_Xe_lines}
\setlength{\tabcolsep}{1.05mm}
\begin{tabular}[t]{rclcrclclcc}
\hline\noalign{\smallskip}
\multicolumn{3}{c}{experimental } && \multicolumn{3}{c}{relative} && explanation              && theoretical\\
\multicolumn{3}{c}{energy}        && \multicolumn{3}{c}{intensity}&&                          && energy\\
\multicolumn{3}{c}{(eV)}          && \multicolumn{3}{c}{(\%)}     &&                          && (keV)  \\
\hline\noalign{\smallskip}
4560   &$\pm$&  6               && $\approx$1250 &$\pm$ &  45           && L X--rays               && \\
6262   &$\pm$& 19                       &&  29.9 &$\pm$ & 3.8           && $\rm{\bar{p}Xe}$(29--28)&& 6219\\
6960   &$\pm$& 14                       &&  44.5 &$\pm$ & 3.3           && $\rm{\bar{p}Xe}$(28--27)&& 6928\\
7771   &$\pm$& 15                       &&  22.4 &$\pm$ & 2.4           && $\rm{\bar{p}Xe}$(27--26)&& 7750\\
8680   &$\pm$& 20                       &&  10.3 &$\pm$ & 1.7           && $\rm{\bar{p}Xe}$(26--25)&& 8704\\
9877   &$\pm$& 19                       &&  16.8 &$\pm$ & 1.6           && $\rm{\bar{p}Xe}$(25--24)&& 9822\\
11163  &$\pm$& 10                       &&  20.9 &$\pm$ & 1.4           && $\rm{\bar{p}Xe}$(24--23)&& 11138\\
12717  &$\pm$&  7                       &&  30.2 &$\pm$ & 1.7           && $\rm{\bar{p}Xe}$(23--22)&& 12698\\
14578  &$\pm$&  5                       &&  49.5 &$\pm$ & 2.0           && $\rm{\bar{p}Xe}$(22--21)&& 14563\\
16825  &$\pm$&  4                       &&  62.7 &$\pm$ & 2.1           && $\rm{\bar{p}Xe}$(21--20)&& 16810\\
19564  &$\pm$&  4                       &&  78.0 &$\pm$ & 2.5           && $\rm{\bar{p}Xe}$(20--19)&& 19543\\
22912  &$\pm$&  4                       &&  86.3 &$\pm$ & 2.6           && $\rm{\bar{p}Xe}$(19--18)&& 22901\\
23823  &$\pm$& 24                       &&   9.7 &$\pm$ & 1.1           && $\rm{\bar{p}Xe}$(24--22)&& 23826\\
25828  &$\pm$& 25                       &&  10.6 &$\pm$ & 1.3           &&  \it not identified     && \\
27095  &$\pm$&  4                       &&  \multicolumn{3}{c}{100}     && $\rm{\bar{p}Xe}$(18--17)&& 27073\\
\hline\noalign{\smallskip}
29118  &$\pm$&  8                       && 228.7 &$\pm$ & 7.3           && K X--rays               && \\
\hline\noalign{\smallskip}
30141  &$\pm$& 23                       &&  11.6 &$\pm$ & 1.6           && $\rm{\bar{p}Al}$(7--6)  && 30107\\
31503  &$\pm$& 65                       &&   3.1 &$\pm$ & 1.1           &&\it not identified       && \\
32336  &$\pm$&  4                       &&  91.3 &$\pm$ & 3.0           && $\rm{\bar{p}Xe}$(17--16)&& 32321\\
39040  &$\pm$& 11                       &&  38.7 &$\pm$ & 2.2           && $\rm{\bar{p}Xe}$(16--15)&& 39028\\
42212  &$\pm$& 55                       &&   6.3 &$\pm$ & 1.8           && \it not identified      && \\
47714  &$\pm$& 11                       &&  46.5 &$\pm$ & 2.6           && $\rm{\bar{p}Xe}$(15--14)&& 47708\\
50036  &$\pm$& 61                       &&   6.7 &$\pm$ & 1.8           && $\rm{\bar{p}Al}$(6--5)  && 49989\\
59152  &$\pm$& 17                       &&  45.8 &$\pm$ & 3.3           && $\rm{\bar{p}Xe}$(14--13)&& 59167\\
\hline\noalign{\smallskip}
\end{tabular}
\end{table}

\subsubsection{Electronic K and L X--rays}

The broad structure between the transitions $\rm{\bar{p}}$Xe(18--17) and $\rm{\bar{p}}$Xe(17--16) 
originates from electronic X--rays from partially depleted xenon (Fig.\,\ref{fig:Xe_Kxrays}). 
About two K X--rays are emitted per formed antiprotonic xenon (Table\,\ref{tab:pbar_Xe_lines}). 

The lower limit coincides approximately with electronic states from iodine, which could 
originate from higher lying $\Delta n_{\rm{\bar{p}}}\gg 1$ transitions and require an almost complete 
electron shell. Because of the dominance of $\Delta n_{\rm{\bar{p}}}=1$ 
and 2 Auger transitions, in particular with the outer shells present, the strong decrease of intensity towards 
lower energies seems to be natural. The upper limit is at about the energy of an electronic K$\alpha_{1}$ for 
highly ionised $\rm{\bar{p}}$Xe with significant L--shell ionisation (Table\,\ref{tab:Zeff_Xe}). 
The central energy of 29.1\,keV corresponds to K hole refilling in the presence of a significantly 
populated M shell as can be seen from inspection of the columns with $n_{\rm{\bar{p}}}=30$ in 
Table\,\ref{tab:Zeff_Xe}. Higher and lower energies require a higher and lower degree of ionisation, 
respectively.

No extra intensity from normal xenon K$\alpha$ and K$\beta$ fluorescence could be identified. In 
Fig.\,\ref{fig:Xe_Kxrays}, the arrows mark the positions of the iodine and xenon K$\alpha$ fluorescence lines. 

About 10 L X--rays are emitted per $\rm{\bar{p}}$Xe atom, i.\,e., at least 10 times L--electron emission 
and subsequent refilling (Fig.\,\ref{fig:Xe_Lxrays}). It requires both the presence of M electrons and a 
rather complete depletion of the N shell, because otherwise the L vacancies are filled mainly by radiationless 
processes\,\cite{kes74}. The maximum of the distribution at about 4.5\,keV corresponds to L--X--ray energies 
for depleted N and O shells with commencing M electron emission. The absence of $\rm{\bar{p}}$Xe lines below 
3.5\,keV shows that the M shell does not deplete before $n_{\rm{\bar{p}}}\approx 29$.

\section{Summary}
\label{sec:summary}

The characteristic X--radiation from the antiprotonic noble gases argon, krypton, and xenon has been studied 
in order to elucidade the role of the non--radiative de--excitation mechanism on the X--ray yields of the 
antiprotonic lines. A number of additional lines in the X--ray spectra of the noble gases have been tentatively 
assigned to electronic transitions caused by electron de--excition after Auger emission during the 
antiprotonic cascade. A few lines remain unexplained so far or are not unambiguously assigned. 

In general, the high degree of circularity of the antiprotonic cascade is evident from the specific 
depletion sequence showing up in the spectrum as alternating low and high yield antiprotonic X--ray transitions. 
As concluded earlier\,\cite{bac88} complete ionisation is reached for krypton, which is not the 
case for xenon. Therefore, the $\rm{\bar{p}}$Xe spectrum is mucher richer in electronic L and K X--rays 
than that of $\rm{\bar{p}}$Ar or $\rm{\bar{p}}$Kr. 

Electronic X--rays are concentrated in the energy ranges corresponding to about the emission bands 
of the principal shells. The complexity of states cannot be resolved with semiconductor detectors. 
The details of such structures may be resolved in future by high--resolution devices like crystal spectrometers 
having at least two orders of magnitude higher resolving power. Also Auger electron spectroscopy might be 
resumed being a powerful techniques to explore the initial stages of the de--excitation 
cascade especially if, e.\,g., lowest--density targets like gas jets can be used.

Such studies will become feasible at antiproton facilities like the one planned at the research center 
GSI\,\cite{FLAIR}. There, sufficiently intense low--energy antiproton beams suited for high--statistics antiproton 
stop experiments could be provided.

\section*{Acknowledgements}
Khalid Rashid wishes to express his gratitude to the For\-schungs\-zentrum J\"ulich and the 
University of Kassel for hosting the visits. KR also acknowledges with thanks the grant from 
Alexander--von--Humboldt foundation, Germany, and from the Higher Education Commission, Pakistan. 
The authors are grateful to  K.--P. Wieder of the Forschungs\-zentrum J\"ulich for the numerous ´
drawings and Josef Anton of the University of Kassel for much technical support. Laboratoire 
Kastler Brossel is Unit\'{e} Mixte de Recherche du CNRS, de l'\'{E}cole Normale Sup\'{e}rieure 
et de l'Universit\'{e} Pierre et Marie Curie n$^{\circ}$ C8552.

%

\end{document}